%% file: main.tex
\documentclass[11pt]{article}
\usepackage{graphicx, verbatim}
\usepackage{caption}
\usepackage{subcaption}

\usepackage{hyperref}

\usepackage[ruled, vlined, linesnumbered]{algorithm2e}
\SetKwInput{initialization}{Initialization}
\SetKwInput{iteration}{Iteration}
\ResetInOut{iteration}
\ResetInOut{initialization}

\usepackage{natbib}

\usepackage{float}
\usepackage{amsmath}
\usepackage{bm}
\usepackage{amssymb}
\usepackage[yyyymmdd,hhmmss]{datetime}

\usepackage{dsfont}
\usepackage{color}

\usepackage[textwidth=1.5cm]{todonotes}

\def\({\ensuremath\left(}
\def\[{\ensuremath\left[}
\def\){\ensuremath\right)}
\def\]{\ensuremath\right]}

\DeclareMathOperator{\Var}{Var}

\DeclareMathOperator{\esjd}{ESJD}

\DeclareMathOperator{\ESS}{ESS}
\DeclareMathOperator{\diag}{diag}
\DeclareMathOperator{\corr}{corr}
\DeclareMathOperator{\sign}{sign}

\newcommand{\Li}{l}
\newcommand{\E}[1]{\mathbb{E} \left[ #1\right] }

\newcommand{\Varr}[1]{\Var \left[ #1\right] }

\newcommand{\Normal}{\mathcal{N}}
\newcommand{\RR}{\mathbb{R}}

\newcommand{\testfunc}{\varphi}

\newcommand{\link}{r}

\newcommand{\sumIN}{\sum_{i=1}^N}
\newcommand{\Flow}{\widehat{\Phi}_{\epsilon,L}}
\newcommand{\Flowe}{\Phi}
\newcommand{\Endiff}{\Delta E}

\newcommand{\target}{\gamma}
\newcommand{\targetnt}{\pi_t}
\newcommand{\targetn}{\pi}
\newcommand{\targett}{\target_t}
\newcommand{\targettmo}{\target_{t-1}}
\newcommand{\targetntmo}{\targetn_{t-1}}

\newcommand{\Vartarget}[1]{\Var_{\targetn} #1 }

\newcommand{\targetspace}{\RR^d}
\newcommand{\normtwo}[1]{ \left\| #1 \right\|_2^2 }
\newcommand{\normtworoot}[1]{ \left\| #1 \right\|_2 }
\newcommand{\normM}[1]{ \left\| #1 \right\|_M^2 }

\newcommand{\denspot}{\target}
\newcommand{\denskin}{f}

\newcommand{\upot}{\mathcal{L}}

\newcommand{\joint}{\mu}

\newcommand{\propkernelt}{\mathcal{K}_t}

\newcommand{\unifzo}{\mathcal{U}[0,1]}

\newcommand{\mom}{p}
\newcommand{\pos}{\particle}

\newcommand{\momt}{\mom_t}

\newcommand{\particle}{x}
\newcommand{\particlet}{\particle_t}

\newcommand{\particleti}{\particle_t^i}
\newcommand{\particletmo}{\particle_{t-1}}

\newcommand{\particletmores}{\tilde{\particle}_{t-1}}
\newcommand{\particleres}{\tilde{\particle}}

\newcommand{\setparticleti}{\left\{\particlet^i \right\}_{i \in 1:N}} 
\newcommand{\setparticletires}{\left\{\tilde{\particle}_t^i \right\}_{i \in 1:N}}

\newcommand{\setparticletmoires}{\left\{\tilde{\particle}_{t-1}^i \right\}_{i \in 1:N}} 

\newcommand{\setparticlewti}{\left\{\particleti, \weightnti\right\}_{i \in 1:N}}

\newcommand{\weightnti}{\weightn_t^i}

\newcommand{\weightn}{w}

\newcommand{\amom}{\hat{\mom}}
\newcommand{\apos}{\hat{\particle}}

\newcommand{\mass}{\mathbf{M}}

\newcommand{\mean}{\mu}
\newcommand{\covariance}{\Xi}

\newcommand{\temp}{\lambda}
\newcommand{\tempt}{\temp_t}
\newcommand{\temptmo}{\temp_{t-1}}

\newcommand{\1}[1]{\mathds{1}_{\left\{#1\right\} }}

\newcommand{\prepsilon}{\hat{\epsilon}}
\newcommand{\preL}{\hat{L}}

\begin{document}

\begin{center}
{\bf \Large Adaptive Tuning Of Hamiltonian Monte Carlo Within Sequential Monte Carlo} \vspace{1cm}\\
Alexander Buchholz$^1$, Nicolas Chopin$^2$, Pierre E. Jacob$^3$\\
$^1$ MRC Biostatistics Unit, University of Cambridge, $^2$ ENSAE-CREST, $^3$ Department of Statistics, Harvard University
 \end{center}

\begin{abstract} 
    Sequential Monte Carlo (SMC) samplers form an attractive
    alternative to MCMC for Bayesian computation. However, their performance
    depends strongly on the Markov kernels used to rejuvenate particles.  We
    discuss how to calibrate automatically (using the current particles)
    Hamiltonian Monte Carlo kernels within SMC. To do so, we build upon the
    adaptive SMC approach of \cite{fearnhead2013adaptive}, and we also suggest
    alternative methods. We illustrate the advantages of using HMC kernels within an SMC sampler
    via an extensive numerical study. 
\end{abstract}


\input{intro}

\input{smc}

\input{hmc}

\input{tuning}


\input{Experiments}
\input{discussion}


\bibliography{my_library}{}
\bibliographystyle{apalike}

\end{document}

%% file: intro.tex
\section{Introduction} 

Sequential Monte Carlo (SMC) samplers
\citep{neal2001annealed,chopin_ibis,del2006sequential} approximate a target
distribution $\pi$ by sampling particles from an initial distribution
$\targetn_0$, and moving them through a sequence of distributions $\targetnt$
which ends at $\pi_T=\pi$.  In Bayesian computation this approach has several
advantages over Markov chain Monte Carlo (MCMC).  In particular, it enables the
estimation of normalizing constants and can thus be used for model choice
\citep{zhou2016toward}. Moreover, particles can be propagated mostly in
parallel, with direct advantages on parallel computing devices
\citep{murray2016parallel}.  Finally, SMC samplers are more robust to
multimodality, see \citet{schweizer2012nonmulti} and \cite{jasra2015error}. 

SMC samplers iterate over a sequence of resampling, propagation and reweighting
steps. The propagation of the particles commonly relies on MCMC kernels, that
depend oftentimes on some tuning parameters. Choosing these
parameters in a sensible manner is challenging and has been of interest both
from a theoretical and practical point of view; see
\cite{fearnhead2013adaptive, schafer2013sequential, beskos2016convergence}.

One type of MCMC kernels that has raised attention recently is HMC (Hamiltonian
Monte Carlo).  HMC has originally been developed in Physics
\citep{duane1987hybrid}, and introduced to the Statistics community by
\citet{neal1993bayesian}.  It has become a standard MCMC tool for sampling
distributions with continuously differentiable density functions on
$\targetspace$ \citep{neal2011mcmc}. The main appeal of HMC is its better
mixing (compared to, say, Metropolis samplers) in high-dimensional problems
\citep{beskos2013optimal,mangoubi2017rapid}. 

This paper compares methods for the automatic tuning of HMC
kernels within SMC. A few previous papers and thesis considered using HMC kernels within
SMC \citep{gunawan2018subsampling, burda2018consumer, daviet2018inference,kostov2016}, but
without focusing on tuning the kernels. In our experience, properly
tuning MCMC kernels within SMC has a big impact on performance, and
particularly so for HMC kernels. As a matter of fact, calibration of HMC kernels
is recognised as a challenging problem in the MCMC literature
\citep[e.g.][]{mohamed2013adaptive, beskos2013optimal,
betancourt2014optimizing, betancourt2016identifying, hoffman2014no}.  The big
advantage of tuning such kernels within SMC is that we have at our disposal a cloud
of particles that inform us on the shape and scale of the current target
distribution. 


We base our approach on the work of
\cite{fearnhead2013adaptive}, which concerned the tuning of generic MCMC
kernels within SMC samplers, and on existing approaches to tuning HMC. 

We apply the proposed SMC sampler with HMC kernels to five examples; three toy
examples, a binary Bayesian regression of dimension up to $95$ and a log
Gaussian Cox model of dimension up to $16,384$.  Our numerical study illustrates
the potential of SMC samplers for inference and model choice in high
dimensions, and the improved performance brought by HMC kernels within SMC
(relative to random walk and Langevin kernels).  We also investigate the
importance of adapting the tempering ladder, and the number of move steps for
diversifying the particles. 

The paper is organized as follows. 
Section \ref{sec:background} reviews SMC samplers and HMC kernels. 
Section \ref{sec:tuning} discusses adaptive tuning procedures for SMC. 
Section \ref{sec:experiments} provides numerical experiments. 
Section \ref{sec:discussion} discusses our results.

%% file: smc.tex
\section{Background} \label{sec:background}

The methods proposed in this article could apply to generic target distributions, but we will focus on posterior distributions
and thus we will use the associated terminology.
We consider the problem of calculating expectations of an integrable test function $\testfunc: \targetspace \mapsto \RR$ 
with respect to a posterior distribution defined as
$
\targetn(\particle) = p(\particle) \Li(y|\particle)/Z.
$
The random variable $\particle$ with density $\targetn(\cdot)$ is 
defined on the space 
$(\targetspace, \mathcal{B}(\targetspace))$, where $\mathcal{B}(\targetspace)$ denotes the Borel set on $\targetspace$. 
Here $p(\particle)$ denotes the prior distribution, $\Li(y|\particle)$ is the likelihood of the observed data
$y$ given the parameter $\particle \in \targetspace$, and $Z=\int_{\targetspace}\Li(y|\particle)p(\particle) d\particle$
denotes the normalizing constant, also called marginal likelihood or evidence. 
We next describe two algorithms widely used to approximate posterior distributions: Sequential Monte Carlo (SMC)
and Hamiltonian Monte Carlo (HMC). These are building blocks for the adaptive SMC procedures discussed in this paper.

\subsection{Sequential Monte Carlo samplers}

\subsubsection{Introducing a sequence of targets: tempering from the prior to the posterior}
Sequential Monte Carlo (SMC) approaches the problem of sampling 
from $\targetn$ by introducing a sequence of intermediate distributions
$\targetn_0, \cdots, \targetnt, \cdots,\targetn_T$ defined on the 
common measurable target space
$(\targetspace, \mathcal{B}(\targetspace))$ for all $t$, and such that 
$\pi_0$ is easy to sample from, and $\pi_T=\pi$. 

We focus on tempering to construct intermediate distributions, that is 
$\targetnt(\particle) \propto p(\particle) \Li(y| \particle)^{\tempt}$, where 
the sequence of exponents $\tempt$ is such that $0 = \temp_0 < \cdots < \tempt <  \cdots < \temp_T = 1$.
These exponents do not need to be pre-specified: they 
may be automatically selected during the run of a SMC sampler, as described later. 
Other choices of sequences of distributions are possible \citep[see
e.g.][]{chopin_ibis, del2006sequential, chopin2013computational}. Note also 
that we assume throughout the article that the prior distribution $p(x)$ is a proper probability distribution,
from which samples can be drawn.

\subsubsection{SMC samplers with MCMC moves on tempered posteriors}

We denote by $\targett(\particle)=p(\particle) \Li(y| \particle)^{\tempt}$ the
unnormalized density associated with $\targetnt(\particle)$, and by $Z_t$ 
the normalizing constant: $Z_t = \int_{\targetspace} \targett(\particle) d \particle$. 
One way of constructing SMC samplers is as follows. 
Suppose that at time $t-1$ an equally weighted particle approximation $\setparticletmoires$ of $\targetntmo$ is available,
with possible duplicates among the particles.
This cloud of particles is then moved with a Markov kernel $\propkernelt$, that leaves the distribution $\targetntmo$
invariant: for each $i$, 
$$
    \particleti \sim \propkernelt(\particletmores^i, d \particle).
$$
Consequently a set of new samples $\setparticleti$ is obtained. 
These particles are then weighted: particle $\particleti$ is assigned 
an importance weight $\weightnti = \targett(\particleti)/\targettmo(\particleti)$, so that the next distribution $\targetnt$
is approximated by the set $\setparticlewti$. After resampling particles according to their weights, one obtains 
again an equally weighted set $\setparticletires$ and the procedure is repeated for the next target distribution
$\targetn_{t+1}$. The resulting algorithm is given in Algorithm \ref{algo:smc_sampler_mcmc}.

Upon completion, the algorithm returns weighted samples 
$\{x_t^i, w_t^i\}_{i\in 1:N}$,
which may be used to estimate expectations with respect to the target
distributions as follows: 
$$
    \frac{\sum_{i=1}^N w_t^i \varphi(x_t^i)}
    {\sum_{i=1}^N w_t^i}
    \rightarrow 
    \mathbb{E}_{\pi_t}\left[\varphi(x) \right]
    $$
as $N\rightarrow +\infty$. The algorithm also returns estimates of the ratios
$Z_t / Z_{t-1}$ (and thus of $Z_T / Z_0$); see line 13. (In the context of tempering, one
may use the path sampling identity to derive an alternate estimate of
$Z_t/Z_{t-1}$, as explained in \cite{zhou2016toward}. However, in our experiments, the
two estimates were very close numerically, so we focus on the former estimate
from now on.) 

The kernels $\propkernelt$ may be constructed, for example, as Metropolis--Hastings kernels
\citep[see e.g.][]{chopin_ibis, jasra2011inference, SimFilStu, fearnhead2013adaptive,zhou2016toward}. 
More details on the general construction of kernels and on optimality  
can be found in \citet{del2006sequential,delsequential_bayescomp}. 
In general Markov kernels may depend on a set of tuning parameters $h$, and are hereafter
denoted by $\propkernelt^h$ to make this dependence explicit, as in Algorithm \ref{algo:smc_sampler_mcmc}. 

\begin{algorithm}[H]
    \KwIn{Number of particles $N$, Initial distribution $\targetn_0$, target distribution $\targetn_T$ and 
    intermediate distributions $\targetn_{t-1}(\particle) \propto p(\particle) \Li(y| \particle)^{\temptmo}$, 
    rule for constructing Markov kernels $\propkernelt^h$ that are $\targetn_{t-1}$ invariant.}

    \KwResult{Set of weighted samples $\setparticlewti$ for $t \in 1:T$ and normalizing 
    constant estimates $\widehat{Z_t/Z_{t-1}}$ for $t \in 1:T$.}
    \initialization{t = 1, $\temp_0 = 0$;}
    \iteration{}
    \While{$\temp_{t-1} < 1$ }{
         \If{$t=1$}{
            \ForEach{$i \in  1:N$}{
            Sample $\particle_1^i \sim \targetn_0$;\\
            }
        }
          \Else{
        Tune Markov kernel parameters $h$ using available particles; see Algorithm \ref{algo:tuning_ft} or \ref{algo:tuning_ours}; \label{algo:line_tune} \\
        \ForEach{$i \in  1:N$}{
            Move particle  $ \particleti \sim \propkernelt^h(\particletmores^i, d \particle)$ ;\\
         }
       (Move step can be iterated for better mixing, see Algorithm \ref{algo:move_steps} \label{algo:line_move} );\\
     }
    Choose the next exponent $\tempt \in (\temp_{t-1},1]$ based on available particles; see Algorithm \ref{algo:choose_temp}; \label{algo:line_choose_temp} \\
        \ForEach{$i \in  1:N$}{
            Weight particle $\weightnti = \frac{\targett(\particleti)}{\targettmo(\particleti)}$ ;\\
         }
         Calculate $\widehat{\frac{Z_t}{Z_{t-1}}} = N^{-1} \sumIN \weightnti $ ;\\
         Resample particles $\setparticlewti$ to obtain $\setparticletires$; \\   
         Set $t = t +1$; \\
     }
     \caption{\label{algo:smc_sampler_mcmc} SMC sampler with MCMC moves on tempered posteriors} 
\end{algorithm}


\subsubsection{Tuning of the SMC sampler} 

Different design choices have to be made for the SMC sampler of Algorithm \ref{algo:smc_sampler_mcmc} 
to be operational.
\begin{description}
\item[(a)] The choice of the next exponent $\lambda_t$, at line \ref{algo:line_choose_temp} of Algorithm \ref{algo:smc_sampler_mcmc}, 
    may be based on available particles; for instance on their effective sample
    size, as explained below. 
\item[(b)] The number of move steps, at line \ref{algo:line_move} of Algorithm \ref{algo:smc_sampler_mcmc}, may be based
    on the observed performance of the Markov kernels; see below. 
\item[(c)] The tuning of the Markov kernel parameters $h$, at line \ref{algo:line_tune} of Algorithm \ref{algo:smc_sampler_mcmc}, 
    may be based on the particles, which is the main difference compared to the usual MCMC setting. The main contribution
    of this paper is to investigate this tuning in the case of HMC kernels, and is described in Section \ref{sec:tuning}.
\end{description}


\paragraph{(a) Choice of the next exponent}

A common approach \citep{jasra2011inference, schafer2013sequential} 
to choose adaptively intermediate distributions within 
SMC is to rely on the ESS \citep[effective sample size, ][]{KongLiuWong}. 
The ESS is a measure of performance for importance sampling estimates \citep{agapiou2017importance}.
This criterion is calculated as follows: 
\begin{eqnarray}
    \ESS(\tempt) = \frac{\left( \sumIN w_t^i \right)^2}{
        \sumIN \left( w_t^i \right)^2}, 
\end{eqnarray}
where $w_t^i = \targett(\particlet^i)/\targettmo(\particlet^i) = \Li(y| \particleti)^{\tempt-\temp_{t-1}}$ 
in the setting considered here.  The ESS is linked to the $\chi^2$-divergence between the distributions $\targetntmo$ and $\targetnt$ or 
equivalently to the variance of the weights. More precisely the ESS is a Monte Carlo approximation of $N/(1+\chi^2(\targetnt, \targetntmo))$ where $\chi^2(\targetnt, \targetntmo)$ is
the $\chi^2$ divergence from $\targetntmo$ to $\targetnt$.

We may choose $\tempt$ by solving (in $\temp$) the equation
$\ESS(\lambda)=\alpha N$, for some user-chosen value $\alpha\in(0,1)$.  The
corresponding algorithm is described in Algorithm \ref{algo:choose_temp}.  The
validity of adaptive SMC samplers based on an ESS criterion is studied in e.g.
\citet{beskos2016convergence}, see also \citet{huggins2015sequential,whiteley2016role}. Another
approach for choosing the sequence of temperature steps is exposed in
\citet{friel2008marginal}. 

\begin{algorithm}[H] \label{algo:choose_temp}
    \KwIn{Target value $\alpha$, likelihood $\Li(y| \particleti)$ for the $N$
    particles $\particleti$, current temperature $\lambda_{t-1}$.}
     \KwResult{Next temperature $\tempt$.}
     Define $\beta^i(\temp) := \Li(y| \particleti)^{\temp-\temp_{t-1}}$ and 
            $$
                \ESS(\temp) = \frac{\left( \sumIN \beta^i(\temp) \right)^2}{
                    \sumIN \left( \beta^i(\temp)  \right)^2};
            $$ 

            \If{$\ESS(1) \geq \alpha N$}{$\lambda_t=1$ }   
            \Else{
            Solve $\ESS(\temp) = \alpha N$ in $\temp\in[\lambda_{t-1}, 1]$,
        using bisection, assign result to $\lambda_t$.}
     \caption{Choice of the next exponent based on the effective sample size.}
\end{algorithm}

\paragraph{(b) Number of move steps}
The mixing of MCMC kernels plays a crucial role in the performance and stability of SMC samplers
\citep[see e.g.][]{del2006sequential,schweizer2012non,ridgway2016computation}. 

For any MCMC kernel targeting a distribution $\pi$, mixing is improved by repeated application of the
kernel, for a cost linear in the number of repetitions. 
We propose to monitor the product of componentwise first-order autocorrelations of the particles to decide how many repetitions to use. 
Autocorrelations are calculated w.r.t. $\setparticletires$, the cloud of particles after 
reweighting and resampling at time $t$. 
After $k$ move steps through the kernel $\propkernelt^h$ the cloud of particles is
$\{\particle_{t,k}^i \}_{i \in 1:N}$. We then calculate the empirical correlation of the component-wise
statistic $\particle^i_{t,k}(j) + \particle^i_{t,k}(j)^2$, where $\particle^i_{t,k}(j)$ denotes
component $j$ of the vector $\particle^i_{t,k}$, using the successive states of the chain $\particle^i_{t,k}(j)$ and $\particle^i_{t,k-1}(j)$. 
This empirical correlation is denoted by $\hat{\rho}_k(j)$. 
This statistic is chosen to reflect the first two moments
of the particles, but is otherwise arbitrary.
We suggest to continue applying the Markov kernel until a large fraction (e.g. $90\%$) of the product of the first order
autocorrelations drops below a threshold $\alpha'=0.1$, for example.
The resulting algorithm is described in Algorithm \ref{algo:move_steps}. 

\begin{algorithm}[H] \label{algo:move_steps}
    \KwIn{Particles $\setparticletires$, proposal kernel $\propkernelt^h$.} 
    \KwResult{Particles after $k$ move steps $\{\particle_{t,k}^i \}_{i \in 1:N}$.}
    \initialization{$\{\particle_{t,0}^i \}_{i \in 1:N} \leftarrow \setparticletires$, $k\leftarrow 0$.}
    \While{ $\# \{j: \prod_{l=1}^k \hat{\rho}_{l}(j) >\alpha'\}/d \geq 10 \%$} {
        Set $k \leftarrow k+1$; \\
        Move particle $\particle^i_{t,k} \sim \propkernelt^h(\particle^i_{t,k-1}, d\particle)$ for all $i$; \\
        Calculate the correlation $\hat\rho_{k}(j)$ for all $j$; 
    }
     \caption{Adaptive move step based on autocorrelations.}
\end{algorithm}

Instead of using component-wise autocorrelations,
one could also draw on recent work on the performance evaluation of MCMC algorithms
in multidimensional spaces \citep{vats2015multivariate} or approaches based on the Stein discrepancy \citep{gorham2015measuring}.

%% file: hmc.tex
\subsection{Hamiltonian Monte Carlo} \label{ssec:hmc}
Hamiltonian Monte Carlo (HMC) consists in proposing moves by solving the 
equations of motion of a particle evolving in a potential. 
We first describe HMC, by following the exposition in \cite{neal2011mcmc},
before turning to existing approaches to the tuning of its algorithmic parameters. 

\subsubsection{MCMC based on Hamiltonian dynamics}
Let $ \upot(\pos) = \log \denspot(\pos)$ be the unnormalized log density of the random variable of interest $\pos$. We introduce 
an auxiliary random variable $\mom \in \RR^d$ with distribution $\mathcal{N}(0, \mass)$, and hence unnormalized log density 
$\log \denskin(\mom) = - 1/2\,\, \mom^T \mass^{-1} \mom$. The joint unnormalized density of $(\mom, \pos)$ 
is given as $\joint(\mom, \pos) = \denskin(\mom)\denspot(\pos)$ and the negative joint log-density is denoted by 
$$ 
    H(\mom, \pos) = -\log \joint(\mom, \pos) = -\upot(\pos)+\frac{1}{2} \mom^T \mass^{-1} \mom. 
$$
The physical analogy of this quantity is the Hamiltonian, 
where the first term denotes the potential energy at position $\pos$, and 
the second term
denotes the kinetic energy of the momentum $\mom$ with the mass matrix $\mass$. 
The movement in time of a particle with position $\pos$ and momentum $\mom$ can be described via
its Hamilton equations, 
$$
   \begin{cases}
    \frac{d \pos}{d \tau} = \frac{\partial H}{\partial \mom} =  \mass^{-1} \mom,  \\
    \frac{d \mom}{d \tau} = - \frac{\partial H}{\partial \pos} = \nabla_\pos \upot(\pos), 
  \end{cases}
$$
where $d \pos /d \tau , d \mom/d \tau$ denote the derivatives of the position
and the momentum with respect to the continuous time $\tau$. The solution of this differential equation 
induces a flow $\Flowe_\tau$ that describes the evolution of a system with initial
momentum and position $(\mom_0, \pos_0)$ such that 
$\Flowe_\tau(\mom_0, \pos_0) = (\mom_\tau, \pos_\tau)$. The solution is (a) energy preserving, 
e.g. $H \left(\mom_\tau, \pos_\tau \right) = H \left( \mom_0, \pos_0  \right)$; (b) 
volume preserving and consequently 
the determinant of the Jacobian of $\Flowe_\tau$ equals one; (c) the flow is reversible w.r.t. time.
In terms of probability distributions 
this means that if $(\mom_0, \pos_0) \sim \joint(\mom, \pos)$ then also 
$(\mom_\tau, \pos_\tau) \sim \joint(\mom, \pos)$. 

In most cases an exact solution of the flow is not available and one has to 
use numerical integration methods instead. One widely used integrator is the 
St\"ormer-Verlet or leapfrog integrator \citep{hairer2003geometric}. The leapfrog integrator 
is volume preserving and reversible but not energy preserving. 
It iterates the following updates:
\begin{eqnarray*}
    \mom_{\tau+\epsilon/2} &=& \mom_{\tau}+\epsilon/2 \nabla_\pos \upot(\pos_\tau), \\
    \pos_{\tau+\epsilon} &=& \pos_{\tau}+\epsilon \mass^{-1} \mom_{\tau+\epsilon/2}, \\
    \mom_{\tau+\epsilon} &=& \mom_{\tau+\epsilon/2}+\epsilon/2 \nabla_\pos \upot(\pos_{\tau+\epsilon}), 
\end{eqnarray*}
where $\epsilon$ denotes the step size of the leapfrog integrator. 
Thus, in order to let the system evolve from $\tau$ to $\tau+\kappa$ with $\kappa = L\times \epsilon$ we need to make $L$ 
steps as described above. 
This induces a numerical flow $\Flow$ such that $\Flow(\mom_\tau, \pos_\tau) = (\amom_{\tau+\kappa}, \apos_{\tau+\kappa})$. 
In general we have $\Endiff_\kappa \neq 0$ where $\Endiff_\kappa = H(\amom_{\tau+\kappa}, \apos_{\tau+\kappa})-H(\mom_\tau, \pos_\tau)$ 
is the variation of the Hamiltonian. 
The dynamics can be used to construct a Markov chain targeting $\mu$ on the joint space, with a Metropolis--Hastings step 
that corrects for the variation in the energy after sampling a random momentum and constructing a 
numerical flow. Algorithm \ref{algo:hmc_kernel} describes the Markov kernel of
HMC. 

\begin{algorithm}[H] \label{algo:hmc_kernel}
     \KwIn{Gradient function $\nabla_\pos \upot(\cdot)$, initial state $\pos_s$, energy function $\Endiff$} 
     \KwResult{Next state of the chain $(\mom_{s+1}, \pos_{s+1})$}
        Sample $\mom_s \sim \mathcal{N}(0_d, \mass)$\\
        Apply the leapfrog integration: $(\amom_{s+1}, \apos_{s+1}) \leftarrow \Flow(\mom_s, \pos_s) $ \\
        Sample $u \sim \unifzo$\\
        \If{$\log(u) \leq \Endiff_s$}{
            Set $\pos_{s+1} \leftarrow \apos_{s+1}$
        }
        \Else{
            Set $\pos_{s+1} \leftarrow \pos_{s}$
        }
     \caption{Hamiltonian Monte Carlo algorithm}
    \end{algorithm}


\subsubsection{Existing approaches to tuning HMC}
The error analysis of geometric integration gives insights that 
allow to find step sizes $\epsilon$ that 
yield stable trajectories.  
For the leapfrog integrator the error of the energy is 
\begin{eqnarray} \label{eq:error_energy}
    \left| H(\amom_{\tau+\kappa}, \apos_{\tau+\kappa}) - H(\mom_\tau, \pos_\tau) \right| \leq C_1 \epsilon^2 ,
\end{eqnarray}
and the error of the position and momentum is
\begin{eqnarray} \label{eq:error_posmom}
    \normtworoot{ \Flow(\amom_{\tau}, \apos_{\tau}) - \Flowe(\mom_{\tau}, \pos_{\tau}) } \leq  C_2 \epsilon^2 ,
\end{eqnarray}
see \citet{leimkuhler2016molecular,bou-rabee_sanz-serna_2018} for more details. 
It can be shown that the constant $C_1>0$ in \eqref{eq:error_energy}
stays stable over exponential long time intervals $\epsilon L \leq \exp(h_0/2\epsilon)$ for some constant $h_0$ 
\citep[Theorem 8.1]{hairer2006geometric}, whereas 
the constant $C_2>0$ in the \eqref{eq:error_posmom} typically grows with $L$. 
Hence, care must be taken when choosing $(\epsilon, L)$.
Using the error control in \eqref{eq:error_energy}, \citet{neal2011mcmc}
following \citet{creutz1988global} provides an informal reasoning
motivating the scaling of $\varepsilon$ as $d^{-1/4}$, 
at least for targets that factorize into products of $d$ independent components.
To maintain a fixed integration time $\varepsilon L$ one should then scale $L$
as $d^{1/4}$.

From a practical point of view the tuning of the HMC kernel 
requires the consideration of the following aspects.
If $\epsilon$ is too large, the numerical integration of the HMC flow becomes unstable 
and results in large variations in the energy and thus a low acceptance rate, see \eqref{eq:error_energy}. 
On the other hand if $\epsilon$ is too small, for a fixed number of 
steps $L$ the trajectories tend to be short 
and high autocorrelations will be observed, see \citet{neal2011mcmc}. 
To counterbalance 
this effect a large $L$ would be needed 
and thus computation time would increase. If $L$ gets too large, 
the trajectories might also 
double back on themselves \citep{hoffman2014no}. 

From a theoretical perspective \citet{beskos2013optimal} and later \citet{betancourt2014optimizing} 
show that the integrator step size $\epsilon$
should be chosen such that acceptance rates 
between $0.651$ and $0.9$ are obtained, when the dimension 
of the target space goes to infinity. 
This idea has been exploited in \citet{hoffman2014no} where 
stochastic approximation is used in order to obtain 
reasonable values of $\epsilon$. 


A different approach is to choose the parameters of the kernel such that the expected squared 
jumping distance (ESJD) of the kernel is maximized, see \citet{pasarica2010adaptively}. 
The ESJD in one dimension is defined as
$$
\esjd = \E{\normtwo{\pos_s - \pos_{s-1}}} = 2(1 - \rho_1) \Vartarget{[\pos]}, 
$$
assuming stationarity.
In this sense maximizing the ESJD of a Markov chain
is equivalent to minimizing the first order autocorrelation $\rho_1$. 
In $d$ dimensions maximizing the ESJD in Euclidean norm amounts to minimizing
the correlation of the $d$ dimensional process in Euclidean norm. 
In the case of HMC this has been discussed by \citet{mohamed2013adaptive} and \citet{hoffman2014no}. 
\citet{mohamed2013adaptive} tune the HMC sampler with Bayesian optimization and 
vanishing adaptation in the spirit of adaptive MCMC algorithms, 
see \citet{andrieu2008tutorial}. The ESJD is then maximized as a 
function of $(\epsilon, L)$. 
\citet{hoffman2014no} discuss the ESJD as a criterion 
for the choice of $L$. As a general idea the simulation of 
the trajectories should be stopped 
when the ESJD starts to decrease. However, this idea has the 
inconvenience of impacting the reversibility of the chain. 
This problem is solved by adjusting the acceptance step in the algorithm. 
The resulting algorithm is called NUTS \citep{hoffman2014no} and is used in 
the probabilistic programming language STAN, see \cite{stan_JSSv076i01}.

\citet{neal2011mcmc} suggests to use preliminary runs in order to find reasonable 
values of $(\epsilon, L)$ and to randomize the values around the chosen values. 
The randomization avoids pathological behavior that might occur 
when $(\epsilon, L)$ are selected badly. 
Other approaches on identifying the optimal trajectory length are discussed in \cite{betancourt2016identifying}. 

Another important tuning parameter is the mass matrix $\mass$, that is used for sampling 
the momentum.  When the target distribution is close to a Gaussian, 
rescaling the target by the Cholesky decomposition of the inverse covariance matrix 
eliminates the correlation of the target and can improve the performance of the sampler. 
Equivalently, the inverse covariance 
matrix can be set to the mass matrix of the momentum. 
This yields the same transformation, see \cite{neal2011mcmc}.
Recently, \cite{girolami2011riemann} suggested to use a position dependent 
mass matrix that takes the local curvature of the target into account. However, the
numerical integrator for the Hamiltonian equation needs to be modified in consequence.

%% file: tuning.tex
\section{Tuning Of Hamiltonian Monte Carlo Within Sequential Monte Carlo} \label{sec:tuning}
We now discuss the tuning of the Markov kernel in 
line \ref{algo:line_tune} of Algorithm \ref{algo:smc_sampler_mcmc}.
The tuning of Markov kernels within SMC samplers can be linked to the tuning of 
MCMC kernels in general. 
One advantage of tuning the kernels 
in the framework of SMC is that information on the intermediate distributions is available 
in form of the particle approximations. 
Moreover, different kernel parameters can be assigned 
to different particles and hence a large number of parameters can be tested 
in parallel. This idea has been exploited by \citet{fearnhead2013adaptive}. We build 
upon their methodology and adjust their approach to the tuning of HMC kernels.

We first describe the tuning of the mass matrix. Second, we present our adaptation of the 
approach of \cite{fearnhead2013adaptive} to the tuning of HMC kernels, abbreviated by FT. Then we
present an alternative approach based on a pre-tuning phase at each intermediate step, abbreviated by \textsc{PR} for preliminary run. Finally, we discuss the 
advantages and drawbacks of the two approaches.

\subsection{Tuning of the mass matrix of the kernels}
The HMC kernels depend on the mass matrix 
$\mass$, used for sampling the momentum. 
Calibrating this matrix based on the particles of the previous iteration
allows to exploit the information generated by the particles. 
In the case of an independent MH proposal 
this has been used by \citet{chopin_ibis} and more recently by \citet{south2019sequential}. 
For the HMC kernel we use the following matrix at iteration $t$ 
\begin{eqnarray} \label{eq:mass_matrix}
    \mass_{t} = \diag(\widehat{\Var}_{\pi_t}[x_t])^{-1}, 
\end{eqnarray}
where $\widehat{\Var}_{\pi_t}[x_t]$ is the particle estimate of the covariance
matrix of target $\pi_t$. 
The restriction to a diagonal matrix 
makes this approach applicable in high dimensions; alternatively a non-diagonal covariance or precision matrix could be estimated
in high dimension by assuming sparsity \citep[see e.g.][]{sparsematrix}. 
Thus, the global structure 
of the target distribution $\targetntmo$ is taken into account and
moves in directions of higher variance are proposed.


\subsection{Adapting the tuning procedure of Fearnhead and Taylor (2013)}
Suppose we are at line \ref{algo:line_tune} during iteration $t$ of Algorithm
\ref{algo:smc_sampler_mcmc}. 
We are now interested in choosing the parameters $h$ of the propagation kernel $\propkernelt^h$. 

\citet{fearnhead2013adaptive} consider the following ESJD (expected
squared jumping distance) criterion: 
\begin{eqnarray} \label{eq:utility}
    g_t(h) = \int \targetn_{t-1}(\particletmo) \propkernelt^h(\particletmo, \particlet) \normM{\particletmo-\particlet} d \particletmo d \particlet
\end{eqnarray}
where $\normM{x - y} = (x-y)^t M^{-1}(x-y)$ stands for the Mahalanobis distance with respect
to matrix $M$; in our case we set $M=\mass_{t-1}$, see \eqref{eq:mass_matrix}. 
(\cite{fearnhead2013adaptive} set $M$ to the covariance matrix of the particles
at time $t-1$, but, again, this requires inverting a full matrix, which may 
be too expensive in high-dimensional problems.) 

By maximizing $g_t(h)$ we minimize the first-order autocorrelation
of the chain. This leads to a reduced asymptotic variance of the chain and
hopefully to a reduced asymptotic variance for estimates obtained from the SMC sampler. 

The tuning procedure referred to as FT has the following steps: 
\begin{enumerate}
    \item Assign different values of $h^i_{t}$ according to their performance 
    to the resampled particles $\particletmores^i$.
    \item Propagate $\particleti \sim \propkernelt^{h^i_{t}}(\particletmores^i, d \particle)$.
    \item Evaluate the performance of $h^i_{t}$ based on $\particleti, \particletmores^i$.
\end{enumerate}

In Step 1, we use the following performance metric, which is a
Rao-Blackwellized estimator of \eqref{eq:utility}:  
\begin{eqnarray} \label{eq:wsjd}
    \tilde{\Lambda}(\particletmores^i, \apos^i_t) = \frac{\normM{ \particletmores^i- \apos^i_t }}{L} \times \min(1, \exp [\Endiff^i_{t}]). 
\end{eqnarray}
Here $\apos^i_t$ is the proposed position based on the Hamiltonian flow $\Flow(\particletmores^i, \mom_t^i)$ before 
the MH step. 
The acceptance rate $\min(1, \exp [\Endiff^i_{t}])$ of the MH step is based on
the variation of the energy $\Endiff^i_{t}$, and serves as weight. 

This metric is the same as in \citet{fearnhead2013adaptive}, except that it is
divided by $L$, in order to account for the fact that the CPU cost of an HMC kernel 
increases linearly with $L$.  


The pairs $h_{t}^i = (\epsilon^i_t, L^i_t)$ are weighted according to the
performance metric $\tilde{\Lambda}(\tilde{x}_{t-1}^i, \hat{x}_t^i)$. 
The next set of parameters $h^i_{t+1}$ are sampled from 
$$
    \chi_{t+1}(h) \propto \sumIN \tilde{\Lambda}(\particletmores^i, \apos^i_t) R(h; h^i_{t}), 
$$
where $R$ is a perturbation kernel.  We suggest to set $R$ to 
$$
R(h; h_{t-1}^i) = \mathcal{TN}(\epsilon; \epsilon^i_{t-1}, 0.015^2) \otimes
\left\{
\frac{1}{3}\1{L_{t-1}^i-1}(L)+\frac{1}{3}\1{L_{t-1}^i}(L)+\frac{1}{3}\1{L_{t-1}^i+1}(L) \right\},
$$
where $\mathcal{TN}$ denotes a normal distribution truncated to $\mathbb{R}^+$. The part in curly brackets corresponds 
to a discrete mixture for the variable $L$.
Thus $\epsilon$ is perturbed by a small (truncated) Gaussian noise, 
and $L$ has an equal chance of increasing, decreasing or staying the same. 
The variance of the Gaussian noise is set to the value used by \cite{fearnhead2013adaptive} (in our 
simulations we found that the tuning was robust to different choices of this value). 
The resulting algorithm is given in Algorithm \ref{algo:tuning_ft}. 

\begin{algorithm}[H] \label{algo:tuning_ft}
     \KwIn{Previous parameters $h^i_{t-1}$, estimator of associated utility $\tilde{\Lambda}(\particleres_{t-2}^i, \apos^i_{t-1})$, $i \in 1:N$, perturbation kernel $R$} 
     \KwResult{Sample of $h^i_{t}=  (\epsilon^i_t, L^i_t)$, $i \in 1:N$}
     \ForEach{$i \in  1:N$}{
        Sample $h_t^i \sim \chi_t(h) \propto \sumIN \tilde{\Lambda}(\particleres_{t-2}^i, \apos^i_{t-1}) R(h; h^i_{t-1})$;
     }
     \caption{(FT) Tuning of the HMC algorithm based on \citet{fearnhead2013adaptive}}
\end{algorithm}


\subsection{Pretuning of the kernel at every time step}
The previous tuning algorithm relies on the assumption that 
parameters suited for the kernel used at time $t-1$ will also achieve 
good performance at time $t$. 

We suggest as an alternative to the previous approach 
the following two-stage procedure: 
\begin{enumerate}
    \item We apply an HMC step (with respect to $\pi_{t-1}$) 
        to the $N$ current particles; for each particle 
        the value of $(\epsilon, L)$ is chosen randomly from a certain 
        uniform distribution (described in next section). 
        We then construct a new random distribution for $(\epsilon, L)$ based on the performance of these 
        $N$ steps (Section \ref{sub:dist_epsL}). 
        The HMC trajectories are then discarded. 
    \item We apply again an HMC step to the $N$ current particles, 
        except this time $(L, \epsilon)$ is generated from the distribution 
        constructed in the previous step. 
\end{enumerate}
We now explain this approach in more detail.

\subsubsection{Range of values for \texorpdfstring{$\epsilon$}{epsilon}}
In the first stage of our pre-tuning parameter, $\epsilon$ is generated from 
$\mathcal{U}[0,\epsilon_{t-1}^\star]$, and $L$ from $\mathcal{U}[0,L_{\max}]$. 
We discuss in this
section how to choose $\epsilon_{t-1}^\star$. The value of the very first $\epsilon_{0}^\star$ is given in Section \ref{sec:experiments}. 

Our approach is motivated by the upper bound in \eqref{eq:error_energy}. 
If for different step sizes $\prepsilon^i_t$ and different momenta and 
positions $(\mom_t^i, \particletmores^i)$ for $i \in \{ 1:N \}$ we observe 
$ \left| \Endiff_t^i \right| = \left| H(\amom_{t}^i, \apos_{t}^i) - H(\mom_t^i, \particletmores^i) \right| $, 
this information may be used to fit a model of the form 
$$\left| \Endiff_t^i \right| = f(\prepsilon^i_t) + \xi_t^i, $$ 
where $\xi_t^i$ is assumed to be such that $\forall i, \E{\xi_t^i}=0, \Varr{\xi_t^i}=\sigma^2 < \infty$ and $f: \RR^+ \rightarrow \RR^+$. 
We may then choose 
$\epsilon^\star$ so that $f(\epsilon^\star) = | \log 0.9 |$. 
This ensures that the acceptance rate of the HMC kernel stays close to $90\%$, following
the suggestions of \cite{betancourt2014optimizing}. 

\begin{figure}[H]
    \centering
    \begin{subfigure}{.45\textwidth}
      \centering
      \includegraphics[width=1\linewidth]{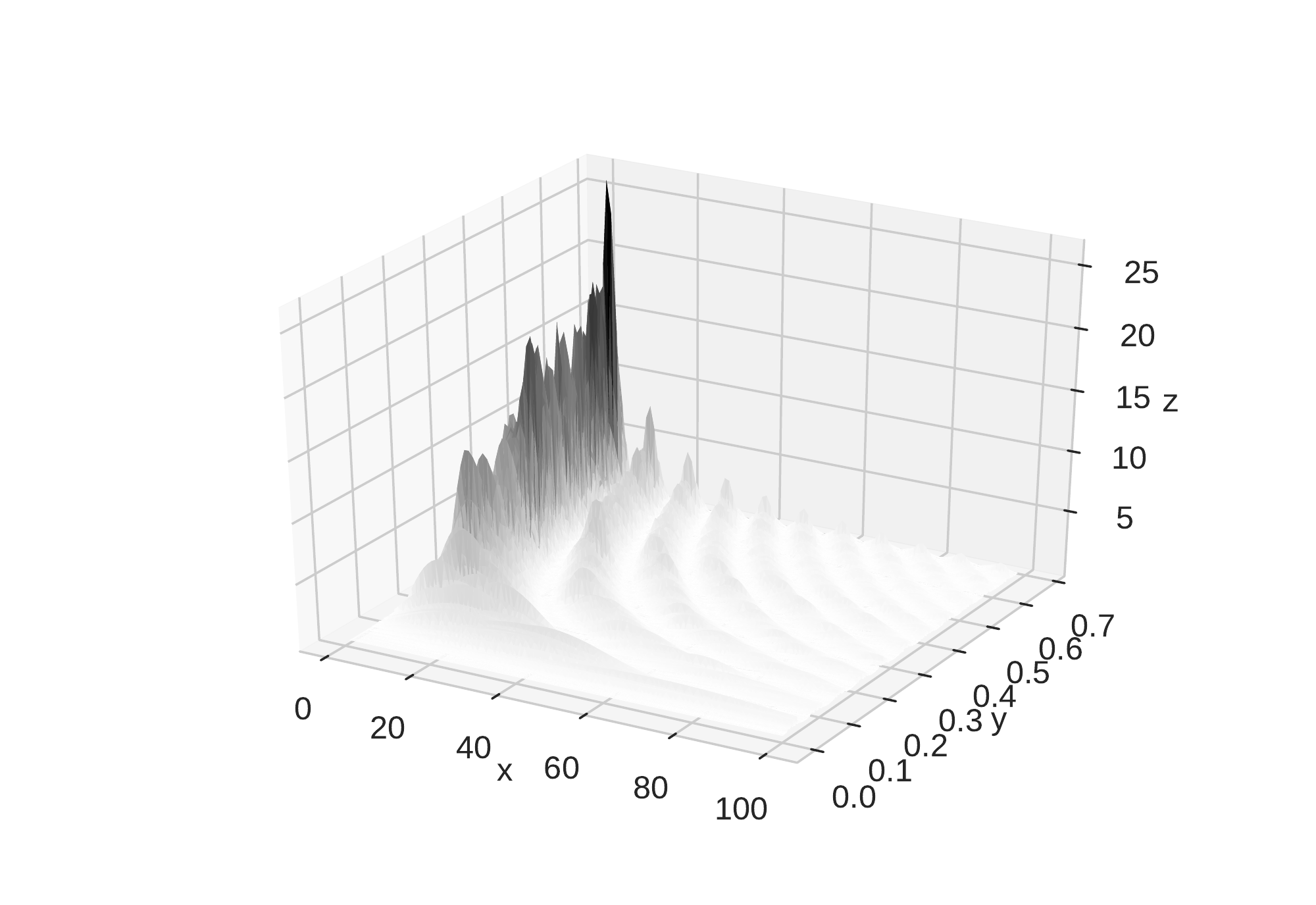} 
    \caption{}
      \label{fig:3D_temp_1}
    \end{subfigure}%
    \begin{subfigure}{.45\textwidth}
      \centering
      \includegraphics[width=1\linewidth]{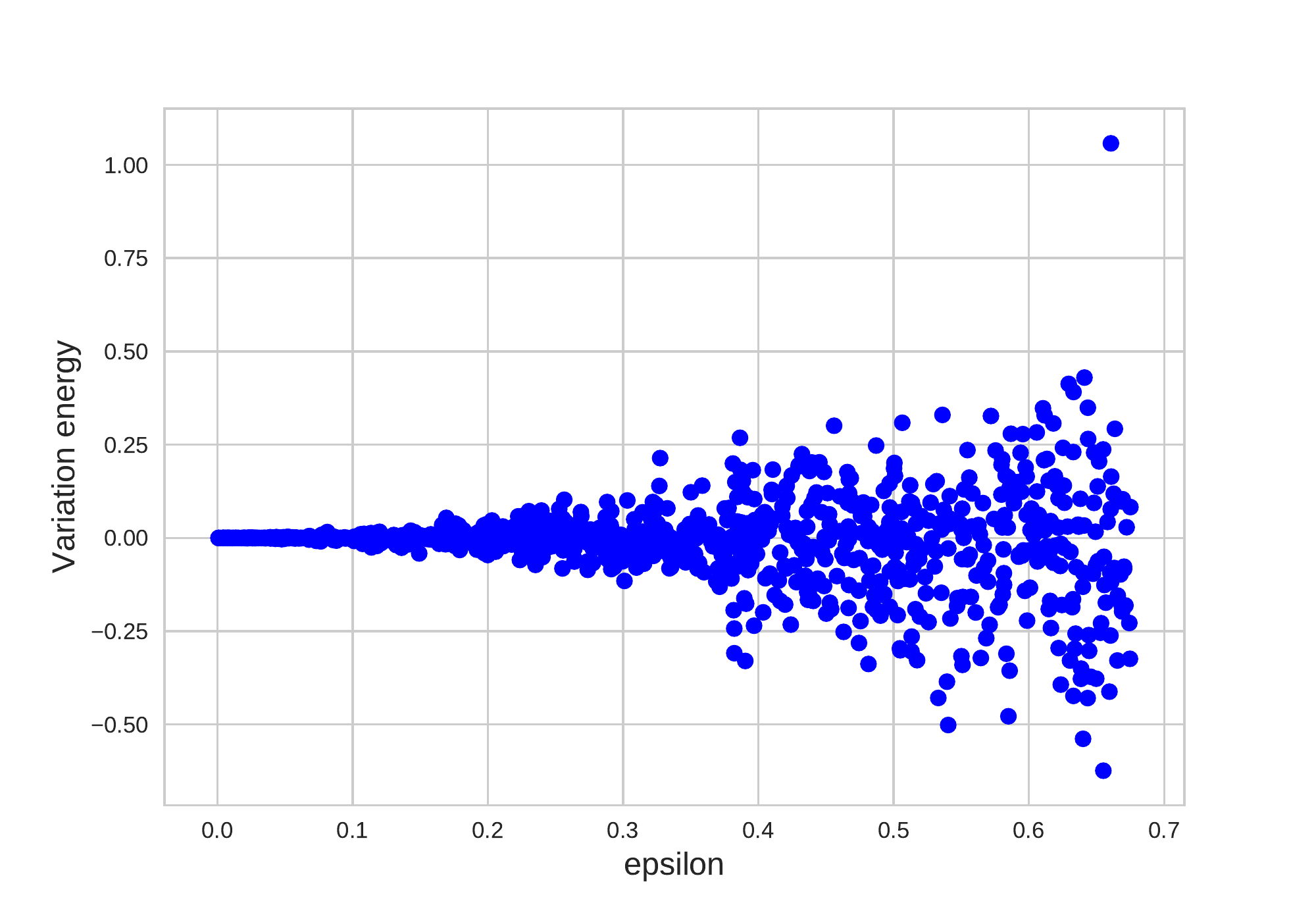} 
    \caption{}
    \label{fig:energy_temp_0}
    \end{subfigure}
    \caption{Tempering of a normal distribution to a shifted and correlated normal distribution in dimension 10 
    (see the example in Section \ref{ssec:normal_to_normal} for more details). 
    Left: The normalized and weighted 
    squared jumping distance (z-axis) as a function of $\epsilon$ (y-axis) and $L$ (x-axis) for the temperature $0.008$. Right:
    Variation of the difference in energy $\Endiff$ as a function of $\epsilon$ for the same temperature. 
    The values of $L$ are randomized.
    Based on an SMC sampler with an HMC kernel based on $N=1,024$ particles.}
\end{figure}

In particular we suggest to use the model 
$$
    f(\prepsilon^i_t)  = \alpha_0 + \alpha_1 (\prepsilon^{i}_t)^2, 
$$
and we minimize the sum of the absolute value errors 
$ \sumIN \left| \xi_t^i\right| $ w.r.t. $(\alpha_0, \alpha_1)$, which amounts to a median
regression. Compared to least squares regression, this approach is more robust to high fluctuations
in the energy which typically occur when $\epsilon$ approaches its stability limit. We illustrate 
this point in Figure \ref{fig:energy_temp_0}.

\subsubsection{Construction of a random distribution for \texorpdfstring{$(\epsilon, L)$}{(epsilon,L)}}
\label{sub:dist_epsL}

Algorithm \ref{algo:tuning_ours} describes how we generate values 
for $(\epsilon, L)$ during the second stage of our pre-tuning procedure. 
In words, these values are sampled from the weighted empirical distribution
that correspond to the $N$ values $(\hat{\epsilon}_t^i, \hat{L}_t^i)$ 
generated (uniformly) during the first stage, with weights given by 
performance metric \eqref{eq:wsjd}. We visualize this metric as a function of $\epsilon, L$ 
in Figure \ref{fig:3D_temp_1}.

\begin{algorithm}[H] \label{algo:tuning_ours}
     \KwIn{Resampled particles $\particletmores^i$, $i \in 1:N$, HMC flow
     $\widehat{\Flowe}_{\cdot, \cdot}$ (targeting $\targetn_{t-1}$), $\epsilon^\star_{t-1}$} 
     \KwResult{Sample of $(\epsilon^i_t, L_t^i)$, $i \in 1:N$, upper bound $\epsilon^\star_{t}$}
     \ForEach{$i \in  1:N$}{
        Sample $\prepsilon^i_t \sim \mathcal{U}[0,\epsilon^\star_{t-1}]$ and 
        $\preL^i_t \sim \mathcal{U}\{1:L_{max}\}$; \\
        Sample $\momt^i \sim \mathcal{N}(0_d, \mass_{t-1})$;\\
        Apply the leapfrog integration: $(\amom^i_t, \apos^i_t) \leftarrow \widehat{\Flowe}_{\prepsilon^i_t, \preL^i_t}(\momt^i, \particleres_{t-1}^i) $;\\
        Calculate $\Endiff^{i}_t$ and $\tilde{\Lambda}(\particleres_{t-1}^i, \apos^i_{t})$
     }
     Calculate $\epsilon^\star_{t}$ based on the quantile regression of $\Endiff^{i}_t$ on $\prepsilon^i_t$ $\forall i \in 1:N$; \\
     Sample $(\epsilon_t^i, L_t^i) \sim \mathcal{C}at\left( w^{i}_t, \{ \prepsilon^i_t, \preL^i_t \}\right)$, 
     where $ w^{i}_t \propto \tilde{\Lambda}(\particleres_{t-1}^i, \apos^i_{t})$ $\forall i \in 1:N$; 
     \caption{(PR) Pre-tuning of the HMC kernel} 
\end{algorithm}

\subsubsection{Range of values for \texorpdfstring{$L$}{L}}

During the first stage of our pre-tuning procedure, $L$ is generated uniformly
within range $[0, L_{\max}]$. The quantity $L_{\max}$ is initialized to some
user-chosen value (we took $L_{\max}=100$ in our simulations). Whenever a large
proportion of the $L_t^i$ generated by Algorithm \ref{algo:tuning_ours} is
close to $L_{\max}$, we increase $L_{\max}$ by a small amount  (5 in our
simulations). Similarly, whenever a large proportion of these values are far
away from $L_{\max}$, we decrease $L_{\max}$ by the same small amount.

\subsection{Discussion of the tuning procedures}

\paragraph{Comparison of the two algorithms}
The difference between the two procedures consists in the pre-tuning phase at each intermediate step of the sampler. 
On one hand, pre-tuning makes the SMC sampler more costly per intermediate step. 
On the other hand this approach makes the sampler more robust 
to a sudden change in the sequence of distributions. 
We illustrate this point in our numerical experiments.

Both of the suggested tuning procedures have computational costs linear in the
number of particles $N$, in line with the other operations performed in the SMC
sampler. 


\paragraph{Other potential approaches to tuning HMC within SMC}

One could try to maximize the squared jumping distance
as a function of the position of every particle, based on 
the associated values of $(\epsilon, L)$. 
However, learning optimal parameters for every position might be challenging,
possibly harder than the original Monte Carlo problem of interest. 
In line with \cite{girolami2011riemann} one could use a position dependent mass matrix that would take 
the geometry of the target space into account, for instance related to higher-order derivatives of the target
probability density function. 

Returning to the choice of $(\epsilon, L)$ one could use an approach based on Bayesian optimization \citep{snoek2012practical}, 
based on the performance of the $(\epsilon_{t-1}^i, L_{t-1}^i)$ at the previous iteration. 
This idea would amount to a parallel version of \cite{mohamed2013adaptive}. 
However, it is not clear how this approach would behave if the underlying distributions evolve over time. 
Not using a pre-tuning step reduces the computational load at the expense of making the sampler potentially less robust.
Moreover, the approach of \cite{fearnhead2013adaptive} already explores the hyperparameter 
space adaptively without requiring additional model specifications.
If framed as a Bandit problem, fixing over time a grid of possible values $(\epsilon, L)$ could be problematic 
if the grid misses relevant parts 
of the hyperparameter space. This holds true in particular 
when using continuous Bayesian optimization, where one typically
has to define some box constraints on the underlying space.

\paragraph{Extensions}


The tuning procedure based on pre-tuning might also be used for tuning random
walk (RW) Metropolis or MALA (Metropolis adjusted Langevin) kernels. In the
first case we may use median regression to find an upper bound for the scale
such that the acceptance rate is close to $23.4 \%$ \citep{roberts1997weak}. In
the second case one may target an acceptance rate of $57.4 \%$
\citep{roberts1998optimscaling}.  (MALA kernels may be viewed as HMC kernels
with $L=1$.) It recently came to our knowledge that the work of
\cite{salomone2018unbiased} also uses a pre-tuning approach for MCMC
kernels within SMC samplers.  A notable difference with our approach is
that \cite{salomone2018unbiased} concentrates on finding a single tuning
parameter, rather than a distribution.

%% file: Experiments.tex
\section{Experiments} \label{sec:experiments} 

Our experiments highlight the importance of adapting SMC samplers, in
particular the parameters of their Markov kernels.  Specifically, we try to
answer the following questions. How important is it to adapt (a) the number of
temperature steps and (b) the number of move steps?  (c) Does our tuning
procedure of HMC kernels provide reasonable values of $(\epsilon, L)$ compared
to other tuning procedures of HMC?  (d) To what extent does HMC within an SMC
sampler scale with the dimension and may be applied to real data applications?
(e) How robust are SMC samplers to multimodality?

For this purpose we compare adaptive (A) and non-adaptive (N) versions 
of HMC-based SMC samplers, where the adaptation may be carried out using 
either the FT approach \citep[our variant of the approach of][]{fearnhead2013adaptive}
or the PR (pre-tuning) approach. We also include in our comparison 
SMC samplers based on random walk (RW) and MALA kernels, and the FT adaptation
procedure. We call our algorithms accordingly: i.e. HMCAFT stands for an SMC
sampler using HMC kernels, which are adapted using the FT procedure. 

In all the considered SMC samplers, the mass matrix $\mass_t$ of the MCMC kernels is 
set to the diagonal of the covariance matrix obtained at the previous iteration. 
Unless otherwise stated, the number of particles is $N=1,024$ and the 
resampling is triggered when the ESS drops below $N/2$. 
The computational load of a given sampler is defined as the number of 
gradient evaluations, plus the number of likelihood evaluations. Note, that this is a conservative choice as
computations of the likelihood and the gradient often involve the same computations. 
Most of our comparisons are in terms of adjusted
variance, or adjusted MSE (mean squared error), by which we mean: variance
(or MSE) times computational load.


All HMC-based samplers are initialized with uniform random draws of $\epsilon$ on $[0,0.1]$ and $L$ on 
$\{1,100\}$. The MALA and RW-based samplers are initialized with random draws of the scale parameter 
on $[0,1]$. The initial mass matrix is set to the identity for all different samplers. 
All samplers choose adaptively the number of move steps based on Algorithm \ref{algo:move_steps}. 
Code for reproducing the results shown below is available under \url{https://github.com/alexanderbuchholz/hsmc}. 

\subsection{Tempering from an isotropic Gaussian to a shifted correlated Gaussian} \label{ssec:normal_to_normal}

As a first toy example we consider a tempering sequence that starts at 
an isotropic Gaussian $\targetn_0 = \Normal(0_d, I_d)$, and finishes at a shifted and correlated Gaussian 
$\targetn_T = \Normal(\mean, \covariance)$, where 
$\mean = 2 \times \mathbf{1}_d$, for different values of $d$. 
For the covariance we set the off-diagonal correlation to $0.7$ and the 
marginal variances to elements of the equally spaced sequence $\tilde{\covariance} = [0.1, \cdots, 10]$.
We get the covariance $\covariance = \diag \tilde{\covariance}^{1/2} \corr(X) \diag \tilde{\covariance}^{1/2}$.
This toy example is rather challenging due to the different length scales 
of the variance, the correlation and the shifted mean of the target. 
In this example the true mean, variance and normalizing constants are available. 
Therefore we report the mean squared error (MSE) of the estimators. 
We use normalized importance weights and thus 
$Z_T/Z_0 = 1$. 

We first compare the following SMC samplers: MALA, HMCAFT and HMCAPR (according to
the denomination laid out in the previous section). We add to the comparison
HMCNFT, an SMC sampler using adaptive (FT based) HMC steps, but where the 
sequence of temperatures is fixed a priori to a long equi-spaced sequence (the
size of which is set according to the number of temperatures chosen adaptively
during one run of HMCAFT). 

\begin{figure}[H]
    \begin{subfigure}{.5\textwidth}
        \centering
        \includegraphics[width=1\linewidth]{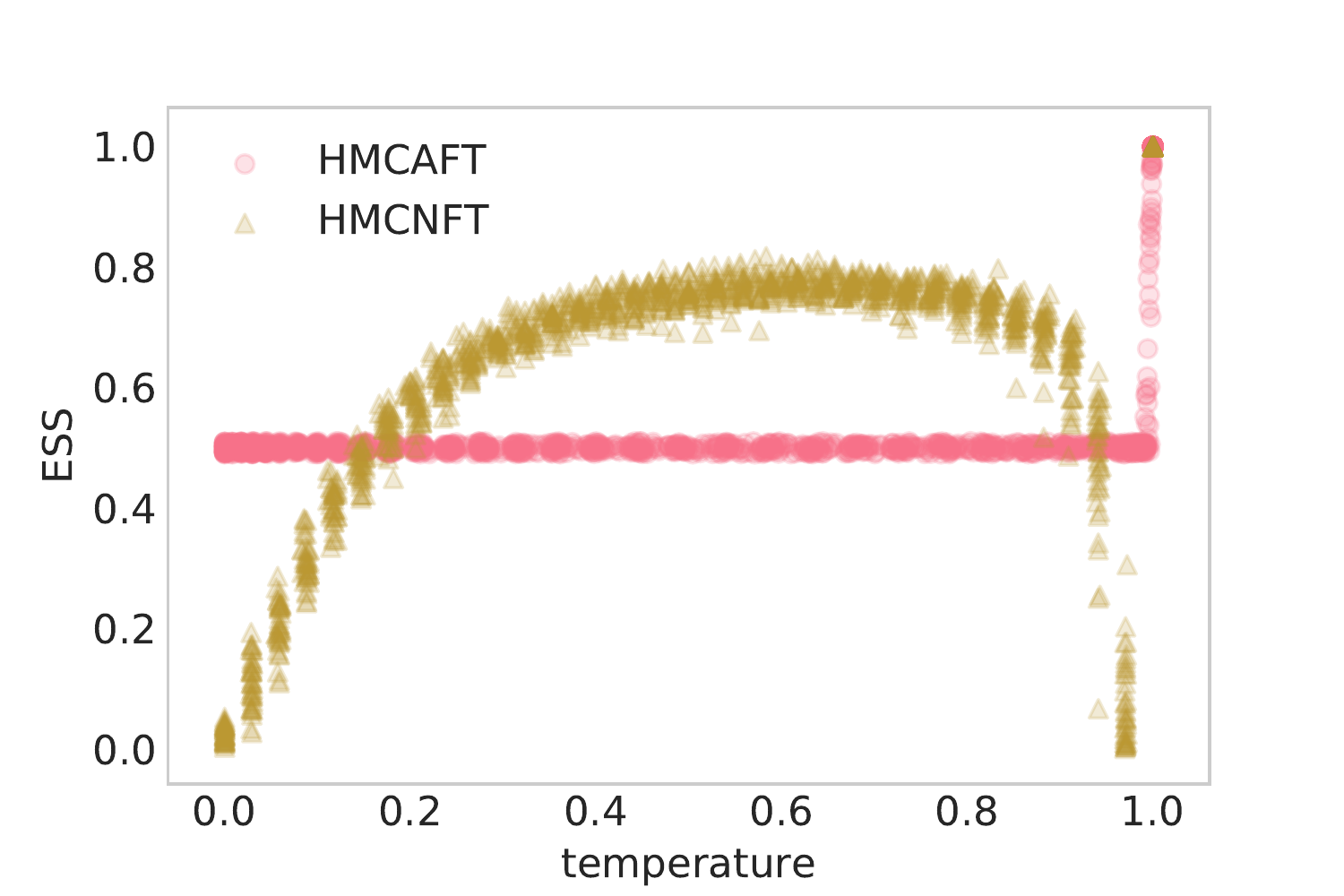}
        \caption{}
        \label{fig:ESS_adaptation_normal_dim_500}
    \end{subfigure}
    \hfill
    \begin{subfigure}{.5\textwidth}
        \centering
        \includegraphics[width=1\linewidth]{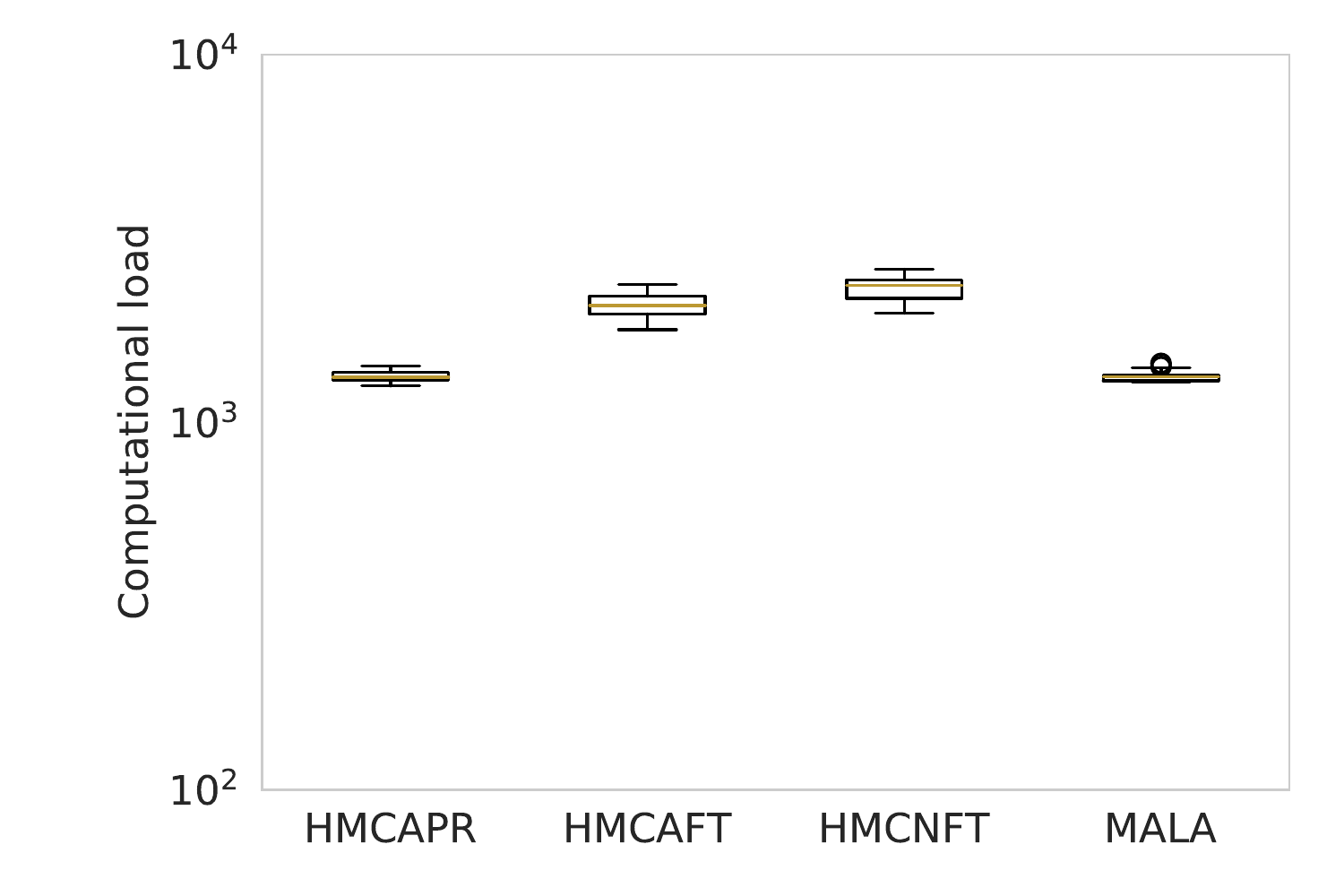}
        \caption{}
        \label{fig:comp_load_normal_dim_500}
    \end{subfigure}
    \caption{ESS and temperature steps in dimension $d=500$ (Figure \ref{fig:ESS_adaptation_normal_dim_500})
    and the computational load of the sampler in terms of the number of total gradient evaluations
    (Figure \ref{fig:comp_load_normal_dim_500}). 
    }
  \end{figure}

Figure \ref{fig:ESS_adaptation_normal_dim_500} plots
the ESS as a function of the temperature, for dimension $d=500$, for algorithms
HMCAFT and HMCNFT. 
Figures 2b and 3 compare the SMC samplers in terms of computational
load (Figure 2b) and adjusted MSE (i.e. MSE times the computational load) for 
the normalizing constant and the expectation of the first component (with 
respect to the target). The results for other components (not shown here) are similar 
to the results for the first component. 

A first observation is that it seems useful to adapt the sequence of
temperatures: HMCNFT is outperformed by all the other algorithms for both
estimates. 
A second observation is that there is no clear ranking between the three other
samplers. HMC-based samplers (and particularly HMCAFT) do perform better than
the MALA-based sampler for the normalizing constant, but the picture is less
clear for the posterior expectation of the first component. It is remarkable 
that, even in dimensions as high as 500, MALA kernels may be 
competitive. 



  \begin{figure}[H]
    \begin{subfigure}{.5\textwidth}
      \centering
      \includegraphics[width=1\linewidth]{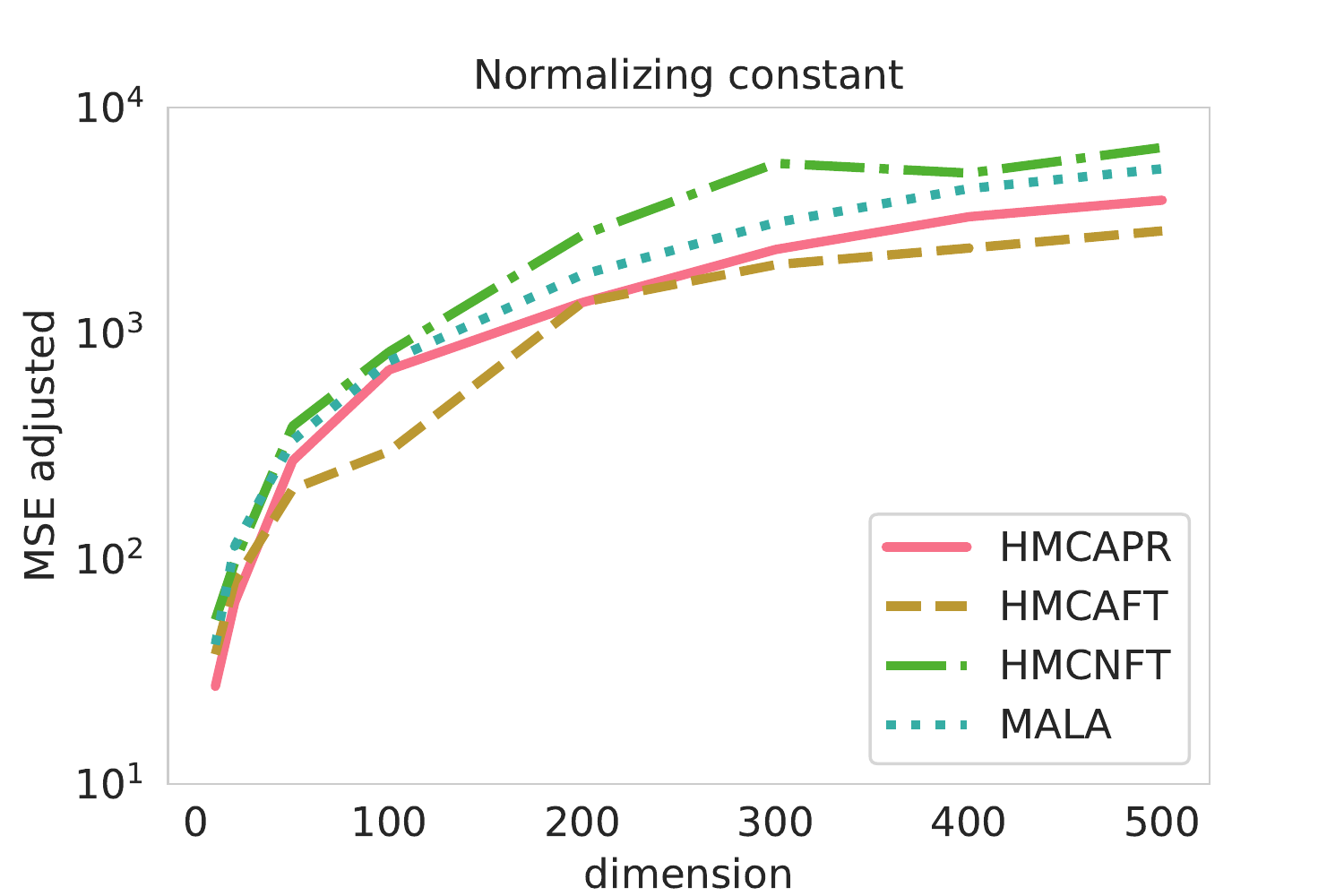}
      \caption{}
    \label{fig:normal_mse_normconst_comp_over_dim}
    \end{subfigure}
    \hfill
    \begin{subfigure}{.5\textwidth}
        \centering
        \includegraphics[width=1\linewidth]{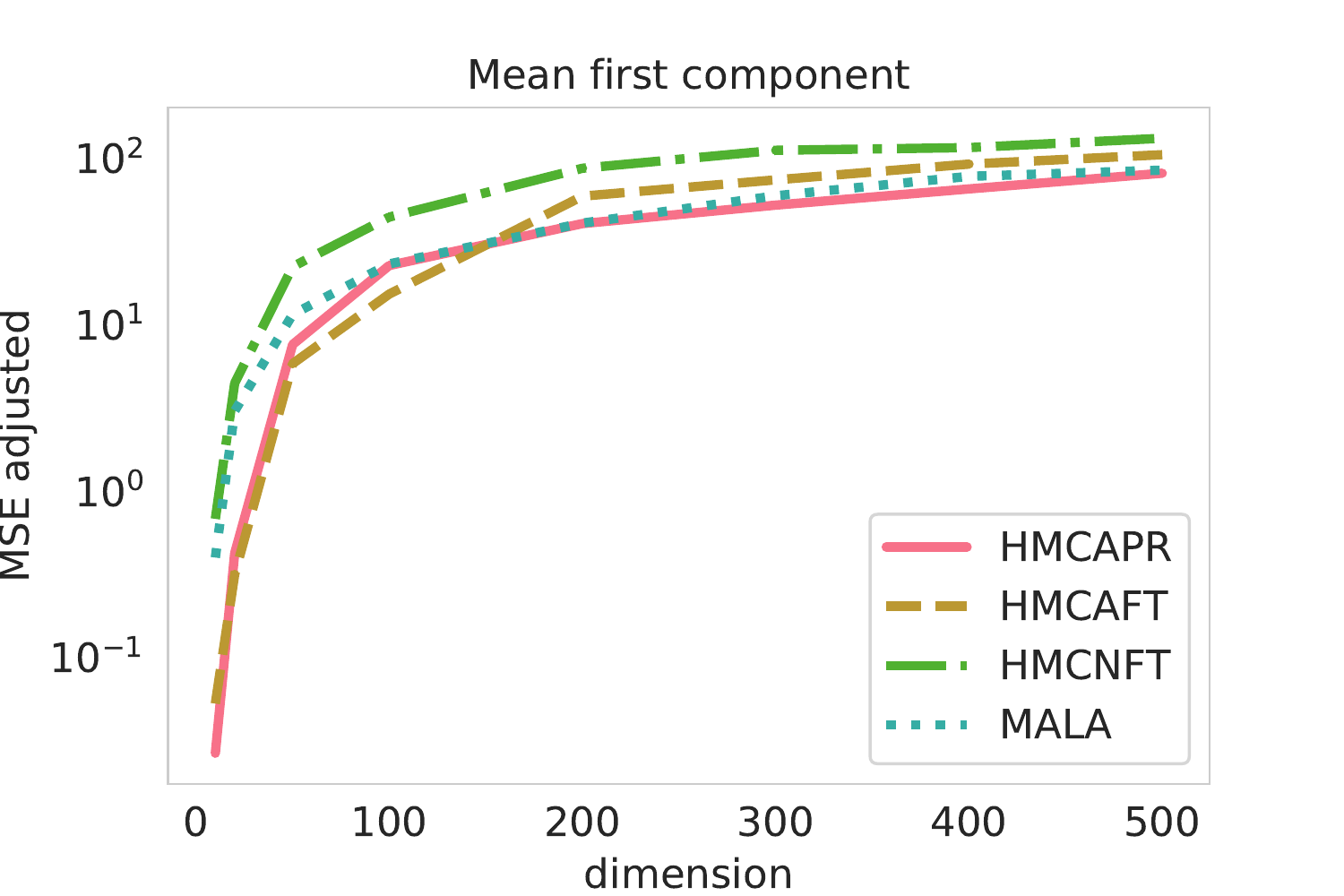}
        \caption{}
        \label{fig:normal_mse_comp1_over_dim}
      \end{subfigure}%
    \caption{
    Mean squared error of the normalization constant (Figure \ref{fig:normal_mse_normconst_comp_over_dim}) 
    and the mean squared error of the first component of the mean (Figure \ref{fig:normal_mse_comp1_over_dim}) 
    multiplied by the computational cost over dimensions. Based on $40$ repetitions of the samplers with $N=1,024$ particles.
    }
  \end{figure}
  
In a second step, we compare the impact of adapting the number of move steps of
the samplers. We restrict the comparison to a MALA-based sampler that uses FT
tuning and an HMC-based sampler that uses PR tuning.  For the non-adaptive
samplers (N) the number of move steps is fixed to a constant number, equal to
the average number of move steps over the complete run of an adaptive sampler.
The temperatures are chosen adaptively. 

\begin{figure}[H]
    \begin{subfigure}{.5\textwidth}
        \centering
        \includegraphics[width=1\linewidth]{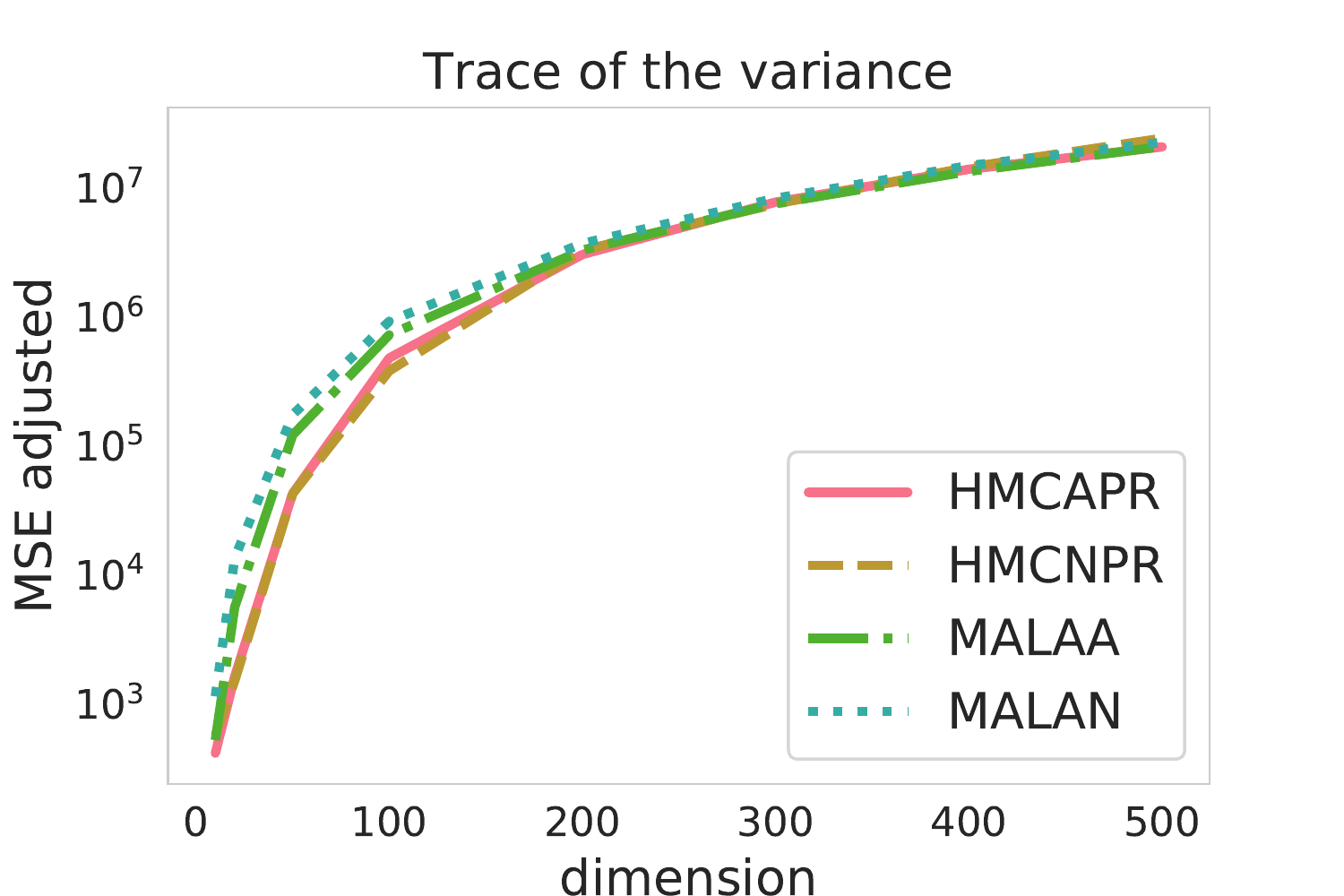}
        \caption{}
        \label{fig:normal_mse_var_comp_over_dim}
    \end{subfigure}
    \hfill
    \begin{subfigure}{.5\textwidth}
        \centering
        \includegraphics[width=1\linewidth]{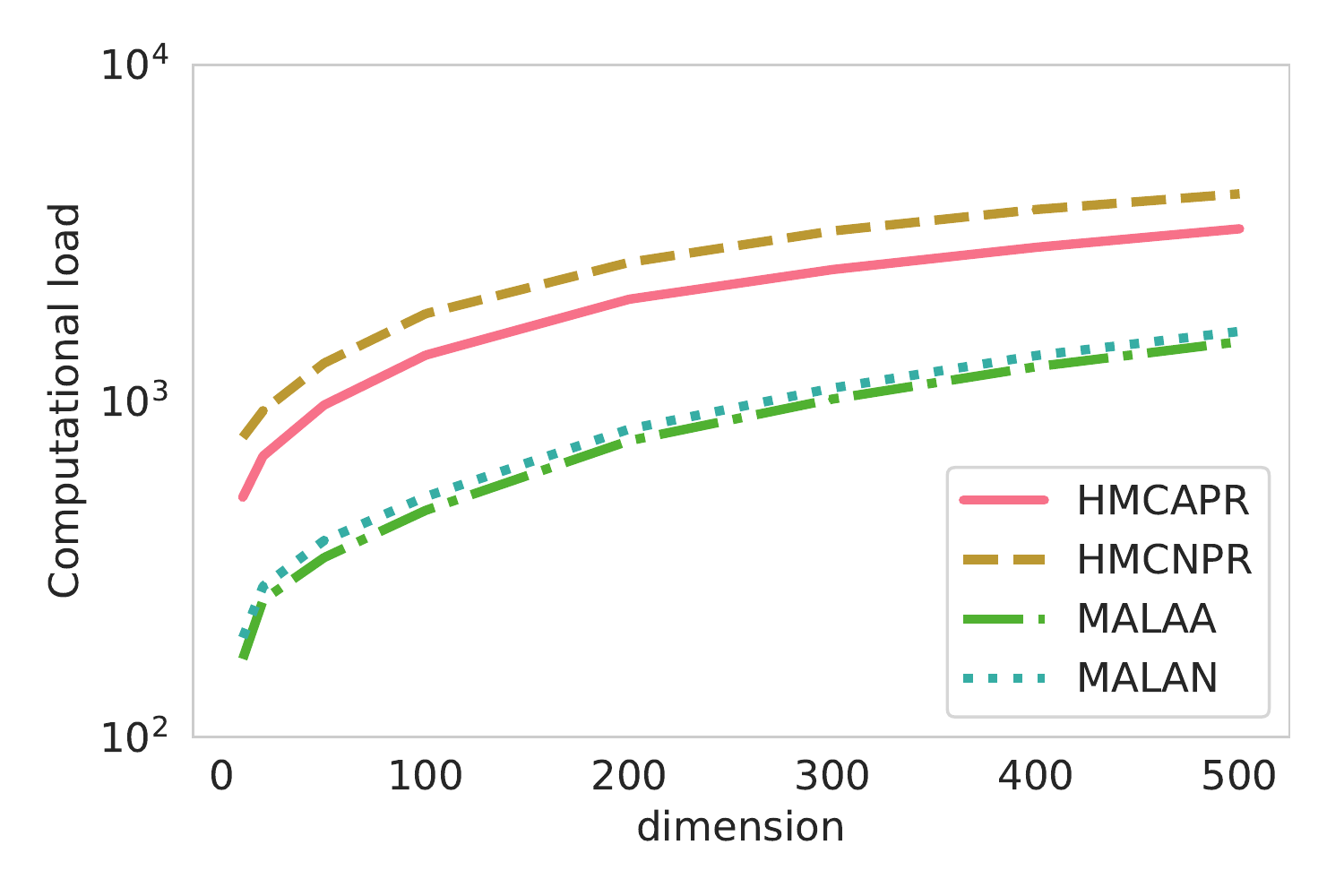}
        \caption{}
        \label{fig:normal_computational_load_over_dim}
    \end{subfigure}
    \caption{Mean squared error of the trace of the estimated variance (Figure \ref{fig:normal_mse_var_comp_over_dim}) 
    multiplied by the computational cost (Figure \ref{fig:normal_computational_load_over_dim}) over dimensions. Based on $40$ repetitions of the samplers with $N=1,024$ particles.
    }
    \label{fig:var_all_normal}
  \end{figure}

We compare the performance of the samplers in estimating the trace of the
variance $\Xi$.  The results are shown in Figure
\ref{fig:normal_mse_var_comp_over_dim} and the associated computational cost is
shown in Figure \ref{fig:normal_computational_load_over_dim}.  The adaptive
samplers seem to offer a similar trade-off in terms of MSE versus computational
load. 
  
In order to assess the performance of the two tuning procedures (FT
and PR), we compare the tuning parameters obtained at the final stage of our
SMC samplers (HMCAFT and HMCAPR) with those obtained from the following MCMC
procedures: NUTS \citep{hoffman2014no} and 
the adaptive MCMC algorithm of \citet{mohamed2013adaptive}. 
NUTS iteratively tunes the mass matrix $\mass$, the number of leapfrog steps $L$ and the 
step size $\epsilon$ in order to achieve a high ESJD (expected squared jumping
distance). 
The adaptive HMC sampler of \citet{mohamed2013adaptive} uses Bayesian optimization in order 
to find values of $(\epsilon, L)$ that yield a high ESJD. 

For this purpose we run our HMC-based SMC samplers and record the achieved ESJD of the HMC kernel at 
the final distribution $\targetn_T$. 
NUTS and the adaptive HMC sampler are run directly on the final target distribution. 
For NUTS we use the implementation available in STAN; for the adaptive HMC sampler 
we reimplemented the procedure of \citet{mohamed2013adaptive}.
After the convergence of the tuning procedures we run the chain for $2,000$ iterations and discard a burnin of $1,000$ samples. 
The ESJD is calculated on the remaining $1,000$ iterations of the chain. 
The results of this comparison are shown in Table \ref{tab:compare_nuts}. 
We see that both the HMC-based SMC samplers, NUTS and the adaptive HMC tuning 
achieve an ESJD of the same order of magnitude. 
Thus, our tuning procedure gives values of $(\epsilon, L)$ that yield an 
ESJD comparable to other existing procedures.


\begin{table}[H]
    \centering
    \begin{tabular}{rrrrrr}
        \hline
            Dimension & SMC HMCAPR &  SMC HMCAFT & SMC MALA & NUTS & adaptive HMC  \\
        \hline
        10 &  50.64 &  61.03 &  9.89 & 59.88 & 134.70 \\
        50 &  174.64 & 255.98 & 30.56 & 204.34 & 190.67 \\
        200 & 639.35 & 989.97 & 85.22 & 1,281.06 & 927.30 \\
        500 & 1,556.27 & 1,311.60 & 154.05 & 2,210.44 & 1,731.04 \\
        \hline 
        \hline
    \end{tabular}
    \caption{Average squared jumping distance of different algorithms for the
        Gaussian target in the first example (based on 40 runs). 
    The results are based on 1,024 samples for the SMC samplers. 
    For the MCMC samplers (NUTS, adaptive HMC) we used a length of  2,000 states
    where the first 1,000 states are discarded as burn-in. }
    \label{tab:compare_nuts}
\end{table}

\subsection{Tempering from a Gaussian to a mixture of two correlated Gaussians}
The aim of our second example is to assess 
the robustness of SMC samplers with respect to multimodality. 
We temper from the prior $\targetn_0 = \Normal(\mean_0, 5 I_d)$ towards a mixture of shifted and correlated Gaussians 
$\targetn_T = 0.3 \Normal(\mean, \covariance_1) + 0.7 \Normal(-\mean, \covariance_2)$, where 
$\mean = 4 \times \mathbf{1}_d$ and we set the off diagonal correlation to $0.7$ for $\covariance_1$ and to $0.1$ for $\covariance_2$. 
The variances are set to elements of the equally spaced sequence $\tilde{\covariance_{j}} = [1, \cdots, 2]$ for $j=1,2$. 
The covariances $\covariance_j$ are based on the same formula as in the first example. 
In order to make the example more challenging we set 
$\mean_0 =  \mathbf{1}_d$. Thus, the starting point of the sampler is slightly biased towards one
of the two modes. 
We evaluate the performance of the samplers by evaluating the signs of the particles and use 
therefore the statistic $T_i :=1/d \sum_{j=1}^d \mathds{1}_{\{\sign X_{j,i} = +1 \}}$, based on a 
single particle.  
We expect a proportion of $30 \%$ of the signs to be positive, i.e. 
$1/N \sumIN T_i \approx 0.3$, if the modes are correctly recovered. 
Our measure of error is based on the squared deviation from this value. 
We consider the following SMC samplers: MALA, RW, HMCAFT, and HMCAPR. 
All the samplers choose adaptively the number of move steps and the temperatures.

\begin{figure}[H]
    \centering
    \begin{subfigure}[t]{.5\textwidth}
      \centering
      \includegraphics[width=1\linewidth]{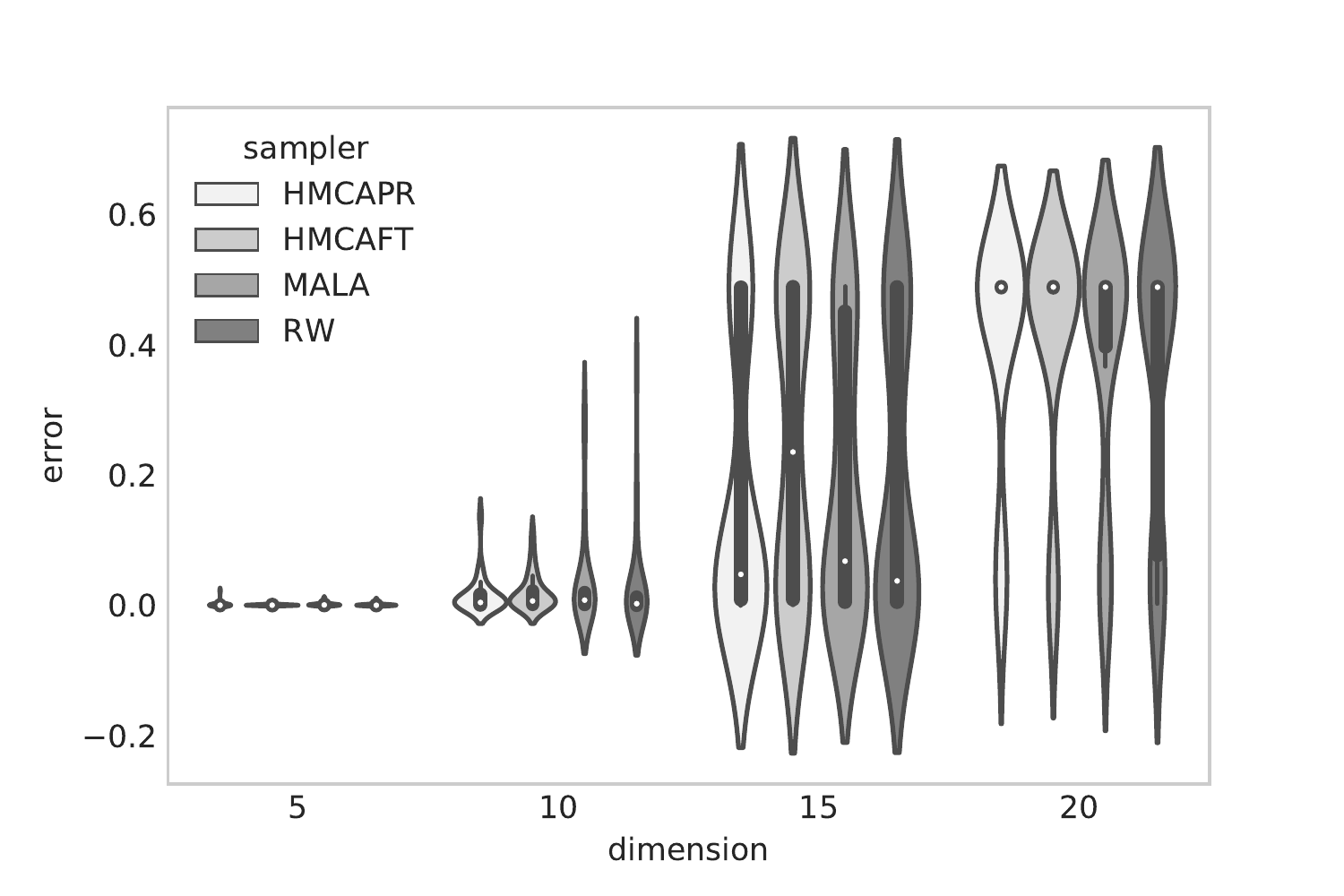}
    \caption{}
      \label{fig:violinplot_mixnormal}
    \end{subfigure}%
    \hfill
    \begin{subfigure}[t]{.5\textwidth}
      \centering
      \includegraphics[width=1\linewidth]{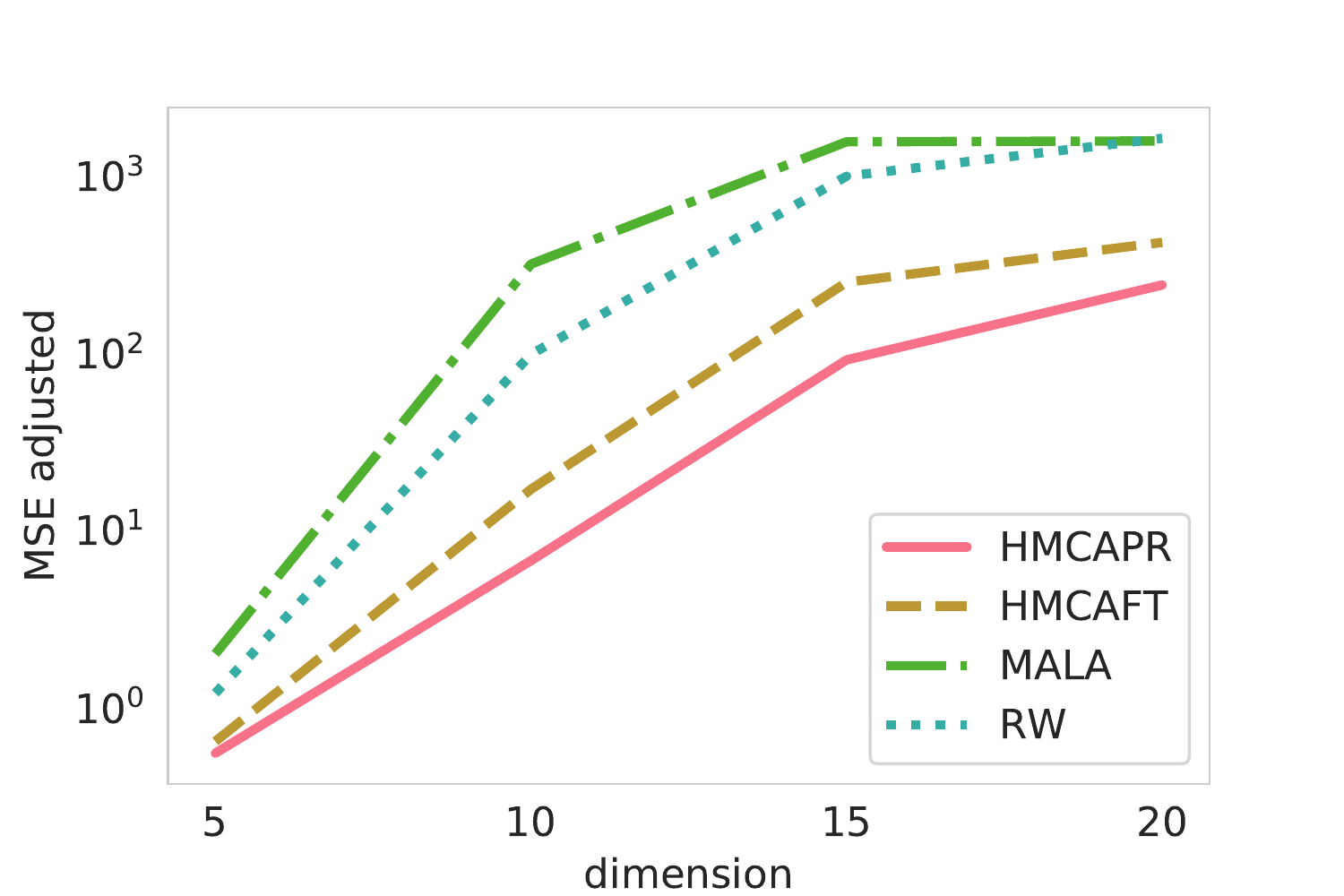}
    \caption{}
    \label{fig:normal_mix_computation_sign}
    \end{subfigure}
    
    \caption{Left: Violinplot of the error, based on the squared difference of $1/N \sumIN T_i - 0.3$. 
      Right: Mean squared error of the proportion of recovered modes adjusted for the computation. 
      Based on 40 runs of the samplers. 
    }
    \label{fig:mix_normal}
\end{figure}

As shown by Figure \ref{fig:normal_mix_computation_sign} the two HMC-based samplers yield a lower error of the recovered modes 
adjusted for computation in moderate dimensions. In terms of the recovery of
the modes all samplers behave comparably,  
see Figure \ref{fig:violinplot_mixnormal}. 
Nevertheless, as the dimension of the problem exceeds $20$ all samplers tend to concentrate on 
one single mode. This problem may be solved by initializing the sampler with 
a wider distribution. However, this approach relies on the knowledge of the location 
of the modes. 

Consequently, SMC samplers are robust to multimodality only in small dimensions and care must be taken
when using SMC samplers in this setting. That said, HMC-based samplers seem
slightly more robust to multimodality; see e.g. results for $d=10$ in Figure 
\ref{fig:violinplot_mixnormal}. See also the recent work of \citet{mangoubi2018does} for the 
performance of HMC versus RWMH on multimodal targets. 

\subsection{Tempering from an isotropic Student distribution to a shifted correlated Student distribution}

As a different challenging toy example we study the tempering from an isotropic 
Student distribution $\targetn_0 = \mathcal{T}_3(0_d, I_d)$ with 
$3$ degrees of freedom 
towards a shifted and correlated Student distribution 
with $\nu = 10$ degrees of freedom, i.e. $\targetn_T = \mathcal{T}_\nu(\mean, \covariance)$. 
The mean and scale matrix are set as in the unimodal Gaussian 
example in \ref{ssec:normal_to_normal}. 

\begin{figure}[H]
    \begin{subfigure}{.5\textwidth}
        \centering
        \includegraphics[width=1\linewidth]{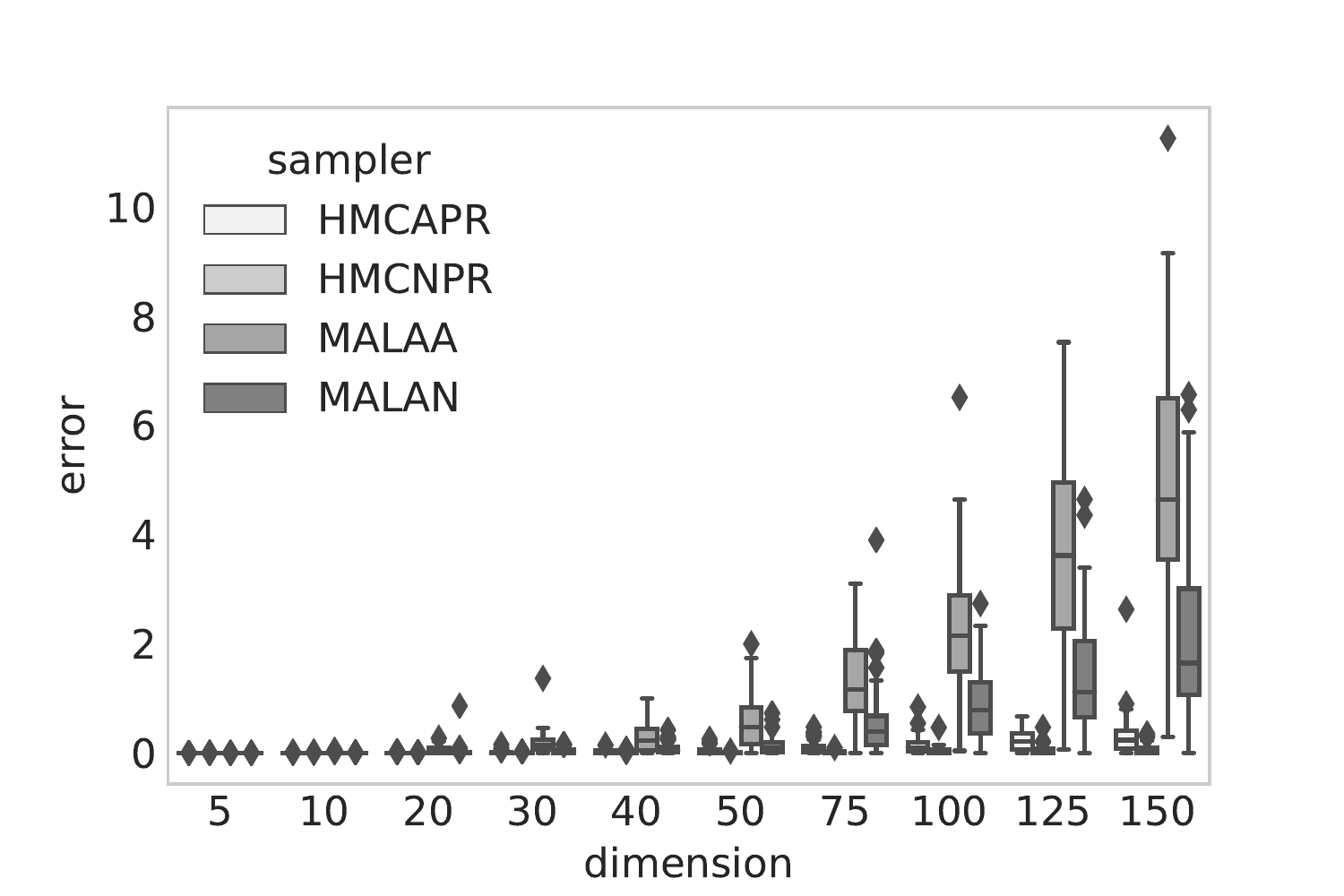}
      \caption{}
        \label{fig:comparison_estimation_normalizing_constant_student}
      \end{subfigure}%
      \hfill
      \begin{subfigure}{.5\textwidth}
        \centering
        \includegraphics[width=1\linewidth]{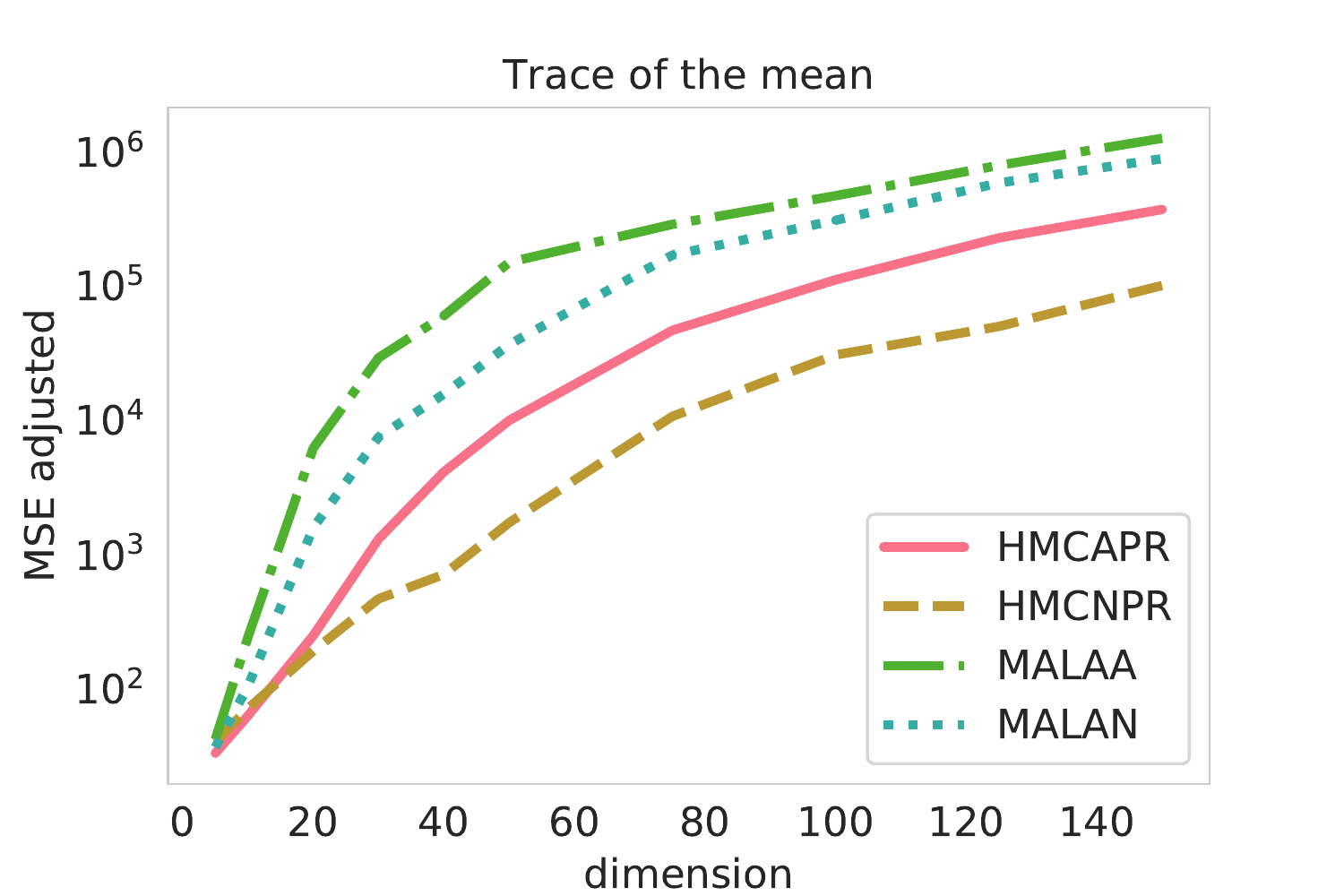}
      \caption{}
        \label{fig:student_mse_comp_over_dim}
      \end{subfigure}%

    \caption{Figure \ref{fig:comparison_estimation_normalizing_constant_student} shows the squared 
    error of the estimator of the normalizing constant. 
    Figure \ref{fig:student_mse_comp_over_dim}
    shows the squared error of the trace of the mean over different dimensions adjusted for computation. 
    The results are based on $100$ runs of the samplers with 
    $N=1,024$ particles.
    }
    \label{fig:mse_all_student}
\end{figure}
This example is more challenging as the target distribution $\targetn_T$ has heavy tails. 
We vary the dimension between $d=5$ and $d=150$. 
The temperature steps are chosen such that an ESS of $90 \%$ is attained. 
The adaptive samplers (A) adjust the number of move steps according to 
Algorithm \ref{algo:move_steps}. 
For the non-adaptive samplers (N)
the number of move steps is fixed to a constant number, 
equal to the 
average number of move steps over the complete run of an adaptive sampler. 
The temperatures are chosen adaptively. The MALA-based sampler uses FT tuning, 
the HMC-based sampler uses our pre-tuning (PR) approach. 

The HMC-based samplers perform better when it comes to estimating the mean 
(see Figure \ref{fig:student_mse_comp_over_dim}) and the normalizing constant 
(see Figure \ref{fig:comparison_estimation_normalizing_constant_student}).
Both samplers based on a MALA kernel tend to work poorly in terms of the estimation 
of the normalizing constant when the dimension increases,
see Figure \ref{fig:comparison_estimation_normalizing_constant_student}. 

In this particular example, we observe that the adaptation of the number of
move steps has some negative impact on the performance of the samplers.  We
suspect that this is due to the heavy tails of the Student distribution: first,
this phenomenon did not occur in our first numerical example, where the target
had light tails (Gaussian); second, \citet{livingstone2016geometric} show that an HMC
kernel cannot be geometrically ergodic when the target is heavy-tailed. Thus,
setting the number of move steps to a fixed, large value may be beneficial for
heavy-tailed targets. Our approach provides a guideline for finding this
number.


\subsection{Binary regression posterior}
We now consider a Bayesian binary regression model. 
We observe $J$ vectors $z_j \in \RR^{d}$ and $J$ outcomes $y_j \in \{0,1 \}$. 
We assume $y_j \sim \mathcal{B}er(\link(z_j^T \beta))$ where 
$\link : \RR \mapsto [0,1]$ is a link function. 
The parameter of interest is $\beta \in \RR^{d}$, endowed with a 
Gaussian prior, i.e. $\beta \sim \Normal(0_d, I_d)$.

When using the logit link function $\link : x \rightarrow \exp(-x)/(1+\exp(-x))$ 
we obtain a Bayesian logistic regression 
where the unnormalized log posterior is given as 
$$
\target(\beta) = \sum_{j=1}^J \left[ (y_j - 1) z_j^T \beta - \log(1+\exp(-z_j^T \beta) \right] - 1/2 \normtwo{\beta}.
$$

When using as link function the cumulative distribution function of a standard normal distribution, denoted $\Phi$, 
one obtains the Bayesian Probit regression.
In this case the unnormalized log posterior is given as
$$
\target(\beta) = \sum_{j=1}^J \left[ y_j \log \Phi(z_j^T \beta) + (1-y_j) \log(1 - \Phi(z_j^T \beta)) \right] - 1/2 \normtwo{\beta}.
$$


We set $\pi_0$ to a Gaussian approximation of the posterior
obtained by Expectation 
Propagation \citep{minka2001expectation} as in \citet{chopin2017leave}. 

We consider two datasets (both available in the UCI repository): 
sonar ($d=61$ covariates, $J=208$ observations) and 
musk ($d=95$, $J=476$, after removing a few covariates that are highly
correlated with the rest). In both cases, an intercept is added, and 
the predictors are normalised (to have mean 0 and variance 1). 

\begin{figure}[H]
    \centering
    \begin{subfigure}{0.5\textwidth}
        \centering
        \includegraphics[width=1\linewidth]{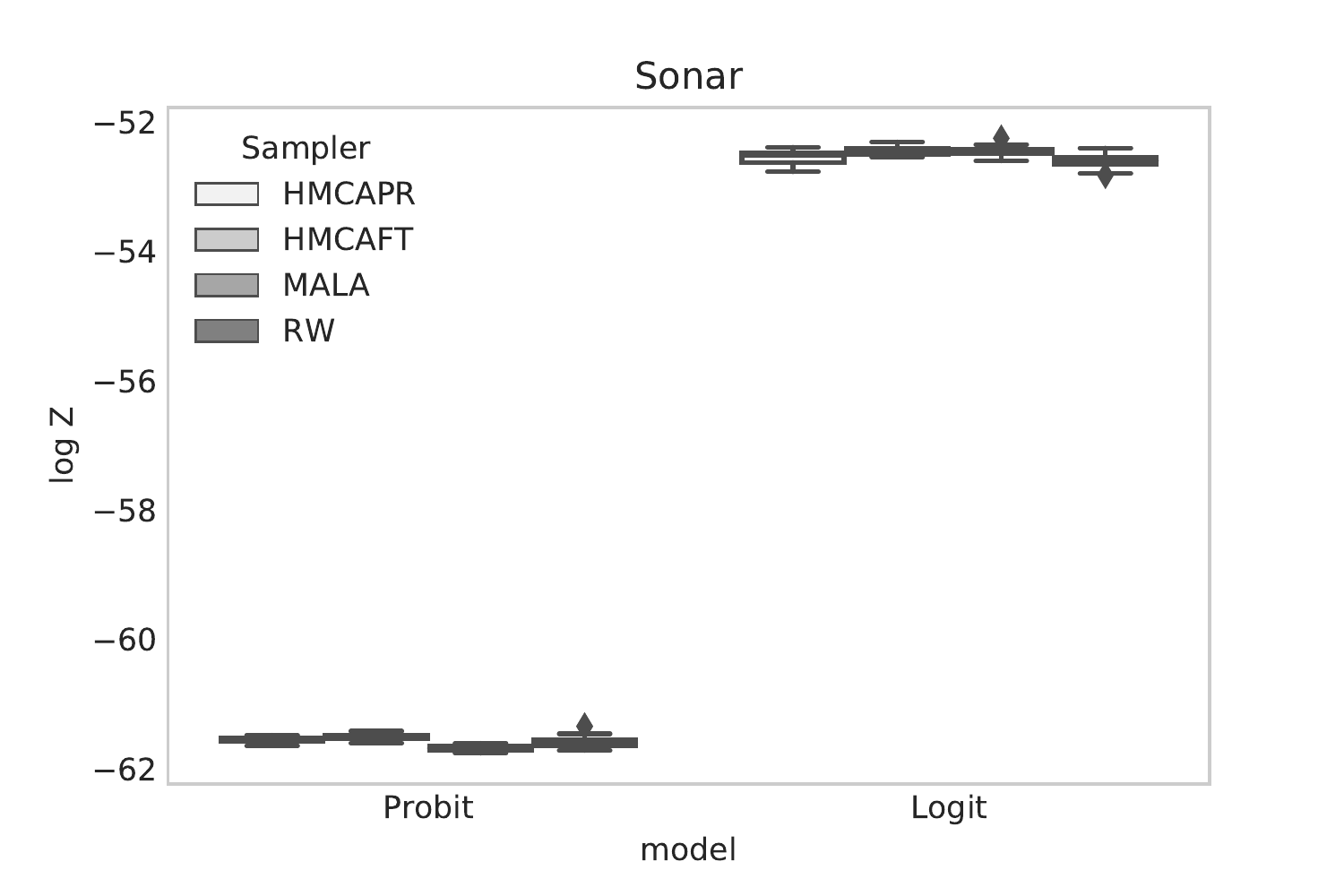}
      \caption{}
        \label{fig:comparison_binary_dim_61_Log_Normconst}
      \end{subfigure}%
      \hfill
      \begin{subfigure}{0.5\textwidth}
        \centering
        \includegraphics[width=1\linewidth]{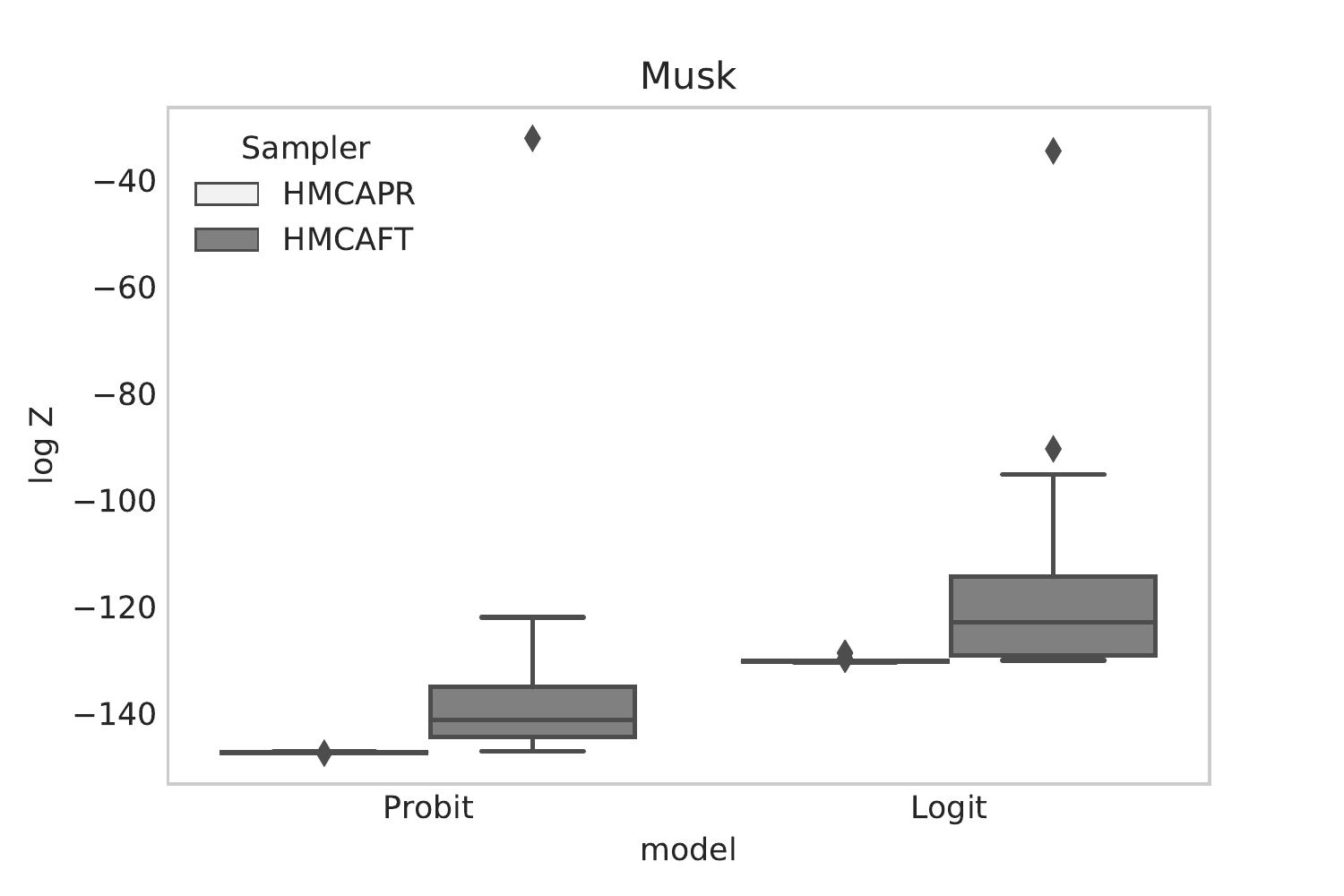}
      \caption{}
      \label{fig:comparison_binary_dim_95_Log_Normconst}
      \end{subfigure}
  
    \caption{Normalization constants obtained for the probit and logit regression based on $40$ runs of the samplers. 
    Figure \ref{fig:comparison_binary_dim_61_Log_Normconst}
    corresponds to the normalization constants obtained for the sonar dataset. The posterior has 61 dimensions. 
    Figure \ref{fig:comparison_binary_dim_95_Log_Normconst} corresponds to the musk dataset.
    The posterior has 95 dimensions.}
    \label{fig:norm_const_binary}
\end{figure}

\begin{figure}[H]
    \centering
    \begin{subfigure}{0.5\textwidth}
        \centering
        \includegraphics[width=1\linewidth]{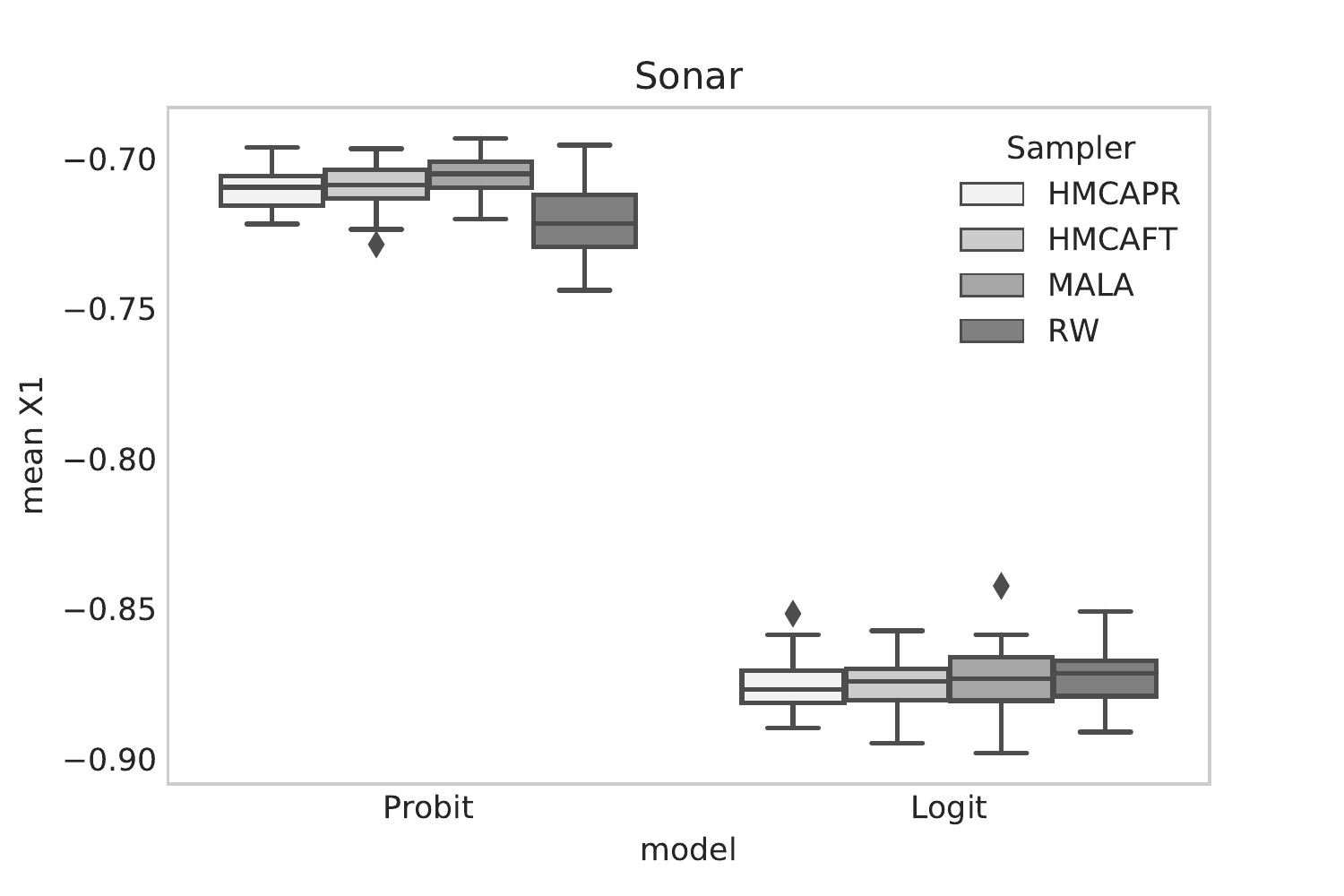}
      \caption{}
        \label{fig:comparison_binary_dim_61_mean_first_comp}
      \end{subfigure}%
      \hfill
      \begin{subfigure}{0.5\textwidth}
        \centering
        \includegraphics[width=1\linewidth]{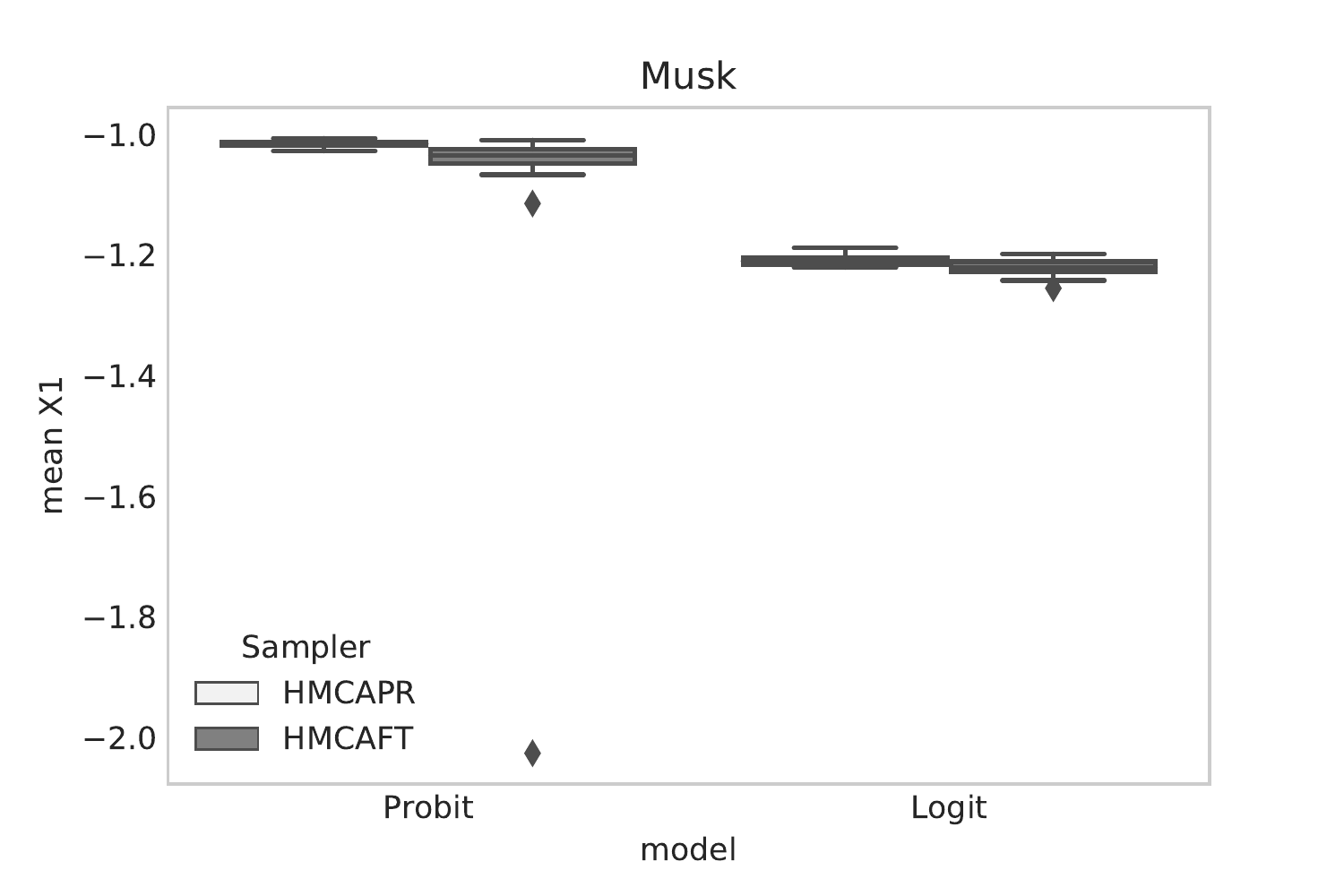}
      \caption{}
      \label{fig:comparison_binary_dim_95_mean_first_comp}
      \end{subfigure}
  
      \caption{Estimated mean of the first component of the posterior obtained for the probit and logit regression.
      Figure \ref{fig:comparison_binary_dim_61_mean_first_comp}
      corresponds to the sonar dataset. 
      Figure \ref{fig:comparison_binary_dim_95_mean_first_comp} corresponds to the musk dataset.}
      \label{fig:mean_predictor_binary}
\end{figure}

We compare the following SMC samplers: RW, MALA, HMCAFT, HMCAPR; see 
Figure \ref{fig:norm_const_binary}
for the estimation of the marginal likelihoods (which may be used to 
perform model choice) and Figure \ref{fig:mean_predictor_binary} for the estimation
of the posterior expectation of the first component. The results for other 
components (not shown here) are similar to the results for the first component. 
For all samplers we adapt the number of move steps and the temperature steps. 

The four samplers perform similarly on on the sonar dataset, while they perform
quite differently on the musk dataset. 
In the latter case, the MALA-based and RW-based samplers did not 
complete after 48 hours of running, so we had to remove them from the comparison. 
(In contrast, the two HMC-based samplers took less than 45 minutes to complete). 
In addition, FT adaptation leads to  a high variance for the normalising
constant. 

Table \ref{tab:binary_regression} reports the logarithm of the adjusted
variances for the four considered samplers, the two considered datasets, the two
considered models (logit and probit) and a varying number of particles for the sonar dataset. 
By and large, HMCAPR seems the most
robust approach; it often gives either the smallest, or a value close to the
smallest, of the adjusted variances. 



\begin{table}[H]
    \centering
    \footnotesize{
    \begin{tabular}{ll|rrrr|rrrr}
        &   &      \multicolumn{8}{c}{Normalizing constant}\\
         &   &      \multicolumn{4}{c|}{Logit} & \multicolumn{4}{|c}{Probit}   \\
        \hline \hline
        Dataset & Dim & HMCAPR & HMCAFT & MALA  & RW    & HMCAPR & HMCAFT & MALA    & RW       \\
        Sonar, $N=2^{10}$ & 61 & 4.183  & \textbf{3.489}  & 5.776 & 6.459 & 3.315  & \textbf{2.98}   & 5.915   & 6.081   \\
        Sonar, $N=2^{13}$& 61  & 2.685  & \textbf{2.233}  & 4.384 & 5.362 & \textbf{2.788}  & 3.686  & 4.883   & 5.285   \\ 
        Musk & 95  & \textbf{6.09}   & 11.519 & -   & -   & \textbf{6.596}  & 8.62   & -     & - \\
        \hline
        &   &      \multicolumn{8}{c}{Mean first component}\\
        &   &      \multicolumn{4}{c|}{Logit} & \multicolumn{4}{|c}{Probit}   \\
       \hline \hline
       Dataset & Dim & HMCAPR & HMCAFT & MALA  & RW    & HMCAPR & HMCAFT & MALA    & RW       \\
       Sonar, $N=2^{10}$ &  61  & -3.409 & \textbf{-3.644} & -0.978 & -0.875 & \textbf{-3.86}  & -3.842 & -1.706  & -0.604    \\
       Sonar, $N=2^{13}$ & 61  & -5.482 & \textbf{-5.792} & -3.668 & -3.255 & \textbf{-5.688} & -3.346 & -3.413  & -2.961   \\ 
       Musk &  95  & \textbf{-1.643} & -1.147 & -  & - & \textbf{-0.633} & 0.014  & -   & -    \\

        \hline
    \end{tabular}
    \caption{Inefficiency of the estimation of the normalizing constant and the mean of the first component
    based on $40$ runs of the different samplers. Smaller is better. 
    The inefficiency is measured as the log of the adjusted variances of the samplers (variance 
    multiplied by the mean number of gradient and likelihood evaluations).
    For the RWMH sampler we adjust by the number of likelihood evaluations only. The best performing sampler for a particular model and number of particles is 
    highlighted in bold. 
    }
    \label{tab:binary_regression}
    }
    
\end{table}

\subsection{Log Gaussian Cox model}

In order to illustrate the advantage of HMC-based SMC samplers in high
dimensions we consider the log Gaussian Cox point
process model in \citet{girolami2011riemann}, applied to the Finnish pine
saplings dataset.  This dataset consists of the geographic position of 126
trees.  The aim  is to recover the latent process
$X \in \RR^{d\times d}$ from the realization of a Poisson process
$Y = (y_{j,k})_{j,k}$ with intensity $m \Lambda(j,k) = m \exp(x_{j,k})$, where
$m = 1/d^2$ and $X = (x_{j,k})_{j,k}$ is a Gaussian process with mean $\E{X} =
\mu \times \mathbf{1}_{d \times d}$ and covariance function
$\Sigma_{(j,k)(j',k')} = \sigma^2 \exp( -\delta(j,j',k,k')/(d \beta) )$, where
$\delta(j,j',k,k') = \sqrt{(j-j')^2+(k-k')^2 }$. 

The posterior density of the model is given as
$$
p(x | y, \mu, \sigma^2, \beta) \propto \left \{  \prod_{j,k}^d \exp( y_{j,k} x_{j,k} 
- m \exp(x_{j,k}) ) \right \} \times \exp \left\{ -1/2 (x-\mu)^T \Sigma^{-1} (x-\mu) \right \}, 
$$
where the second factor is the Gaussian prior density. 

Following the results of \citet{christensen2005scaling} we fix $\beta = 1/33$, $\sigma^2 = 1.91$ and $\mu = \log(126) - \sigma^2/2$. 
We vary the dimension of the problem from $d=100$ to $d=16,384$ by considering different discretizations.
We consider three SMC samplers: MALA, HMCAFT and HMCAPR. The starting
distribution is the prior. 

\begin{figure}[H]
    \centering
    \begin{subfigure}[b]{0.5\textwidth}
      \centering
      \includegraphics[width=\linewidth]{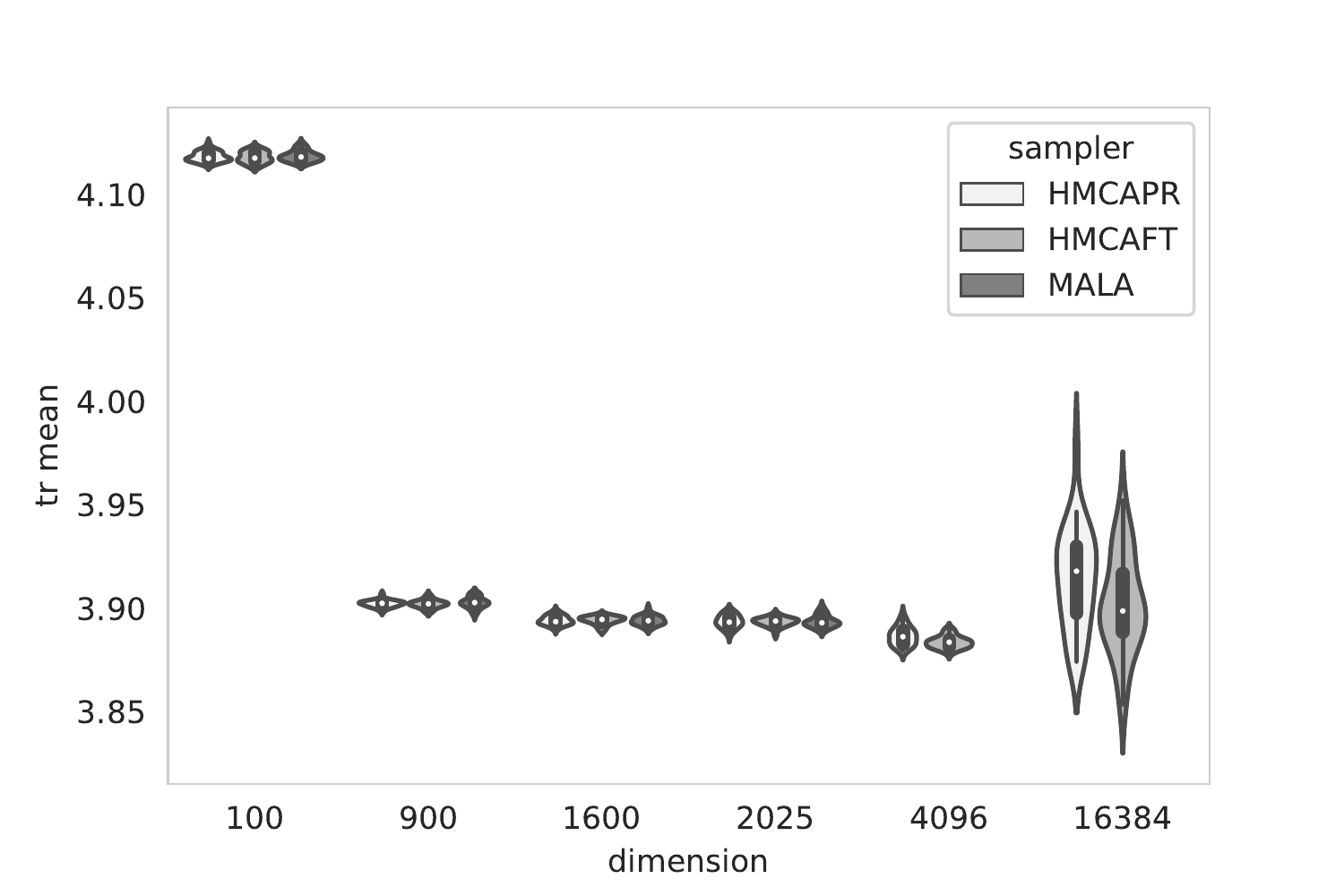}
    \caption{}
      \label{fig:comparison_estimation_trace_mean_log_cox}
    \end{subfigure}%
    \hfill
    \begin{subfigure}[b]{0.5\textwidth}
        \centering
        \includegraphics[width=\linewidth]{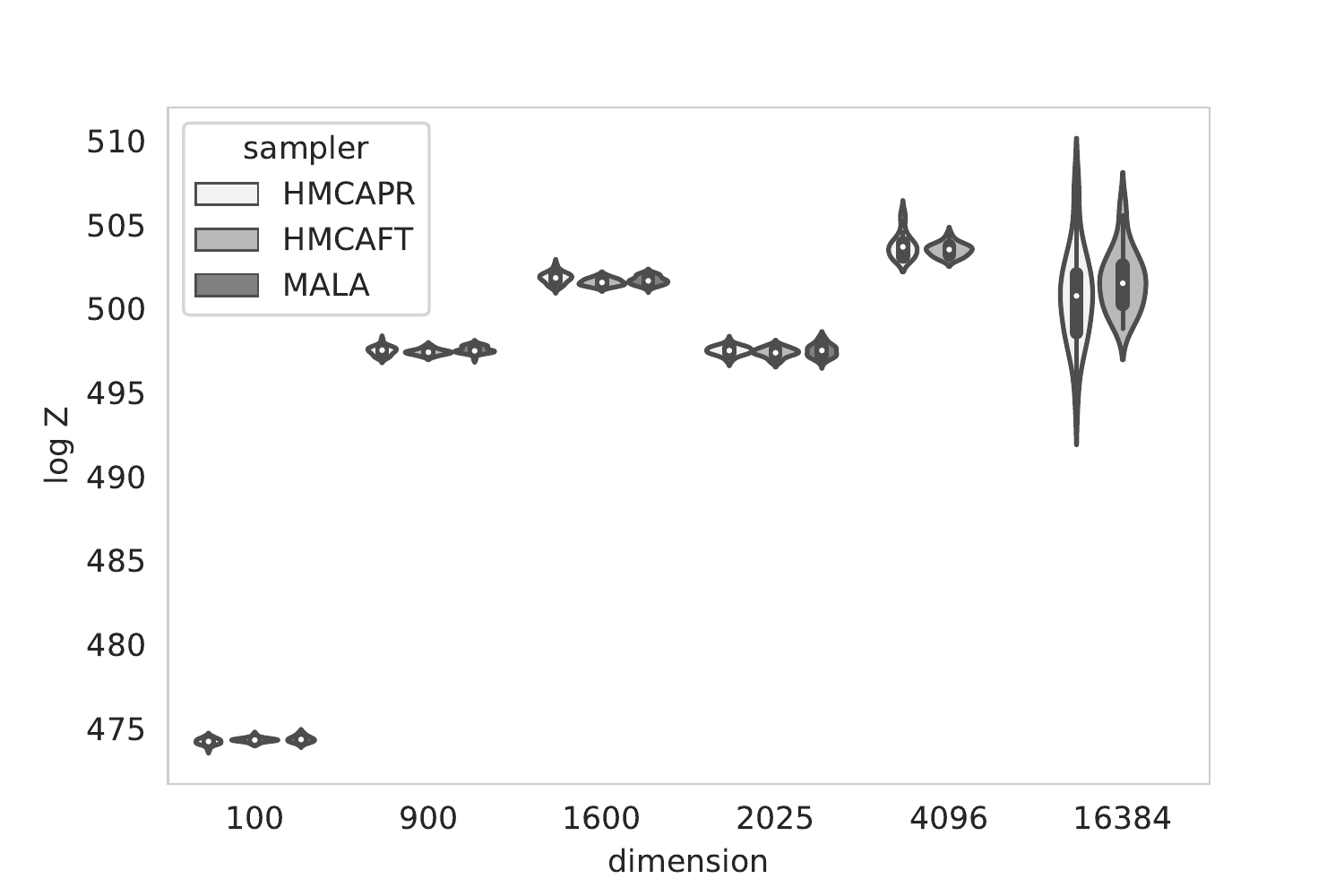}
      \caption{}
        \label{fig:comparison_estimation_normalizing_constant_log_cox}
      \end{subfigure}%
      \hfill
      \begin{subfigure}[b]{\textwidth}
        \centering
        \includegraphics[width=1\linewidth]{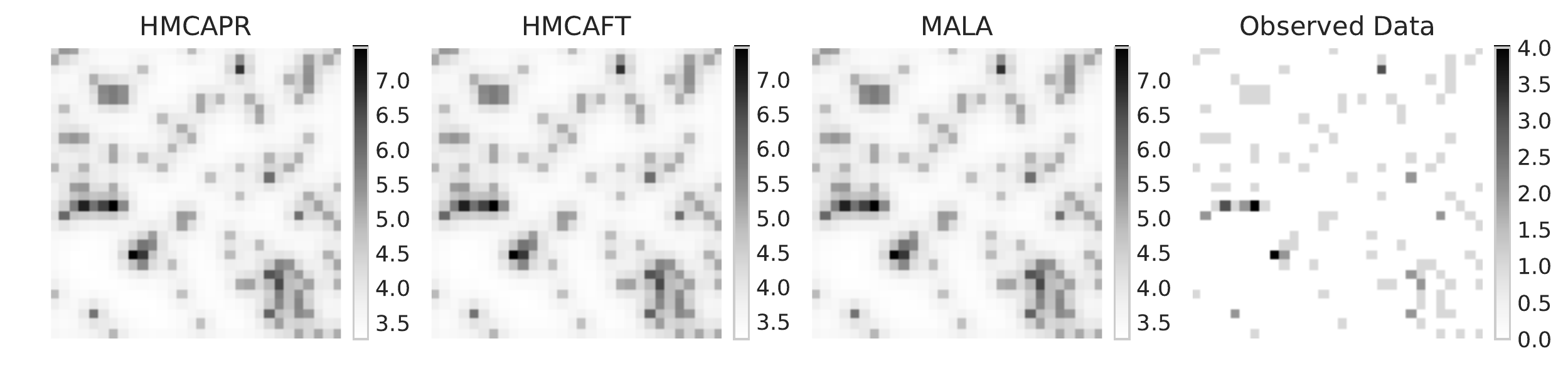}
      \caption{}
      \label{fig:log_cox_model_dim_900}
        \end{subfigure}%

    \caption{Tempering from a normal prior to the posterior of log Gaussian Cox process over various dimensions.
    Figure \ref{fig:comparison_estimation_trace_mean_log_cox} illustrates the estimations of the normalizing constants.
    Figure \ref{fig:comparison_estimation_normalizing_constant_log_cox} illustrates the estimated cumulative posterior mean.
    Figure \ref{fig:log_cox_model_dim_900} 
    illustrates the recovered componentwise posterior mean of the process in dimension $900$. 
    }
\end{figure}

Figure \ref{fig:comparison_estimation_normalizing_constant_log_cox} shows that
the HMC-based samplers estimate well the normalizing constant, even for a large
dimension $d$. 
Moreover, we estimate the cumulative mean throughout different dimensions with
relatively low variance, see Figure
\ref{fig:comparison_estimation_trace_mean_log_cox}. We omitted the simulation
for the MALA-based samplers, as the simulation took excessively long in
dimension $4,096$ due to slow mixing of the kernel.  The estimated posterior
mean of the latent field for dimension $900$ is plotted in Figure
\ref{fig:log_cox_model_dim_900}. 

Table \ref{tab:logcox} reports adjusted variances (variances times
computational load) for the considered SMC samplers. We see that
the MALA-based sampler is not competitive when the dimension increases. 
Regarding the adaptation scheme, FT outperforms PR, especially in high
dimension.


  \begin{table}[H]
    \centering

        \begin{tabular}{l|rrr|rrr} 
            \hline
            &       \multicolumn{3}{c}{Normalizing constant}    & \multicolumn{3}{|c}{Mean first component}    \\
            \hline
            Dim  & HMCAPR & HMCAFT & MALA  & HMCAPR & HMCAFT & MALA  \\
            100   & 2.292  & 2.03   & 2.933 & -6.4   & -5.613 & -5.559 \\
            400   & 3.296  & 2.255  & 3.812 & -5.895 & -5.913 & -4.765 \\
            900   & 3.89   & 2.776  & 4.373 & -6.192 & -6.141 & -4.276 \\
            1,600  & 4.735  & 3.226  & 5.224 & -5.217 & -6.046 & -4.162 \\
            2,500  & 4.5    & 4.072  & 6.246 & -4.405 & -5.636 & -3.476 \\
            4,096  & 7.055  & 5.071  & -  & -3.2   & -4.701 & -   \\
            16,384 & 10.002 & 8.864  & -  & 0.538  & 0.142  & -   \\
            \hline
            \end{tabular} 
    \caption{Inefficiency of the estimation of the normalizing constant and the mean of the first component
    based on $40$ runs of the different samplers. 
   The inefficiency is measured as the log of adjusted variances of the samplers (variance 
   multiplied by the mean number of gradient and likelihood evaluations). Smaller is better. 
    } 
    \label{tab:logcox}
\end{table}

%% file: discussion.tex
\section{Discussion} \label{sec:discussion}

Our experiments indicate that the relative performance of HMC kernels within
SMC depends on the dimension of the problem. For low to medium dimensions, 
RW and MALA are much faster, and tend to perform reasonably well. On the other hand, 
for high dimensions, HMC kernels outperform, sometimes significantly, RW and
MALA kernels. 

The key to good performance of SMC samplers (based on HMC or other kernels) is
to adaptively tune the Markov kernels used in the propagation step.  We have
considered two approaches in this paper.  On posterior distributions with
reasonable correlation our adaption of the approach by
\cite{fearnhead2013adaptive} works best. Our approach based on pre-tuning of
the HMC kernels is more robust to changes in the subsequent target
distributions, as illustrated by  our binary regression example.  This holds in
particular when using SMC samplers for model choice.  Moreover, we showed that,
when sensibly tuned, SMC samplers with HMC kernels can scale to high
dimensional problems.  From a practical point of view and if the structure of
the posterior is unknown the second approach may be the more prudent choice. 




All in all, our methodology provides a principled approach for an automatic adaption of SMC samplers, 
applicable over a range of various models in different dimensions. 

\section*{Acknowledgments}

The first author gratefully acknowledges a GENES research scholarship, a DAAD grant for visiting the third author and funding 
from the EPSRC grant EP/R018561/1. 
The second author is partly supported by Labex Ecodec (anr-11-labx-0047).
The third author gratefully acknowledges support by the National Science Foundation through grants DMS-1712872.

%% file: main.bbl
\begin{thebibliography}{}

\bibitem[Agapiou et~al., 2017]{agapiou2017importance}
Agapiou, S., Papaspiliopoulos, O., Sanz-Alonso, D., Stuart, A., et~al. (2017).
\newblock Importance sampling: Intrinsic dimension and computational cost.
\newblock {\em Statistical Science}, 32(3):405--431.

\bibitem[Andrieu and Thoms, 2008]{andrieu2008tutorial}
Andrieu, C. and Thoms, J. (2008).
\newblock A tutorial on adaptive {MCMC}.
\newblock {\em Statistics and computing}, 18(4):343--373.

\bibitem[Beskos et~al., 2016]{beskos2016convergence}
Beskos, A., Jasra, A., Kantas, N., and Thiery, A. (2016).
\newblock {On the convergence of adaptive sequential Monte Carlo methods}.
\newblock {\em The Annals of Applied Probability}, 26(2):1111--1146.

\bibitem[Beskos et~al., 2013]{beskos2013optimal}
Beskos, A., Pillai, N., Roberts, G., Sanz-Serna, J.-M., and Stuart, A. (2013).
\newblock Optimal tuning of the hybrid {M}onte {C}arlo algorithm.
\newblock {\em Bernoulli}, 19(5A):1501--1534.

\bibitem[Betancourt, 2016]{betancourt2016identifying}
Betancourt, M. (2016).
\newblock {Identifying the optimal integration time in Hamiltonian Monte
  Carlo}.
\newblock {\em arXiv preprint arXiv:1601.00225}.

\bibitem[Betancourt et~al., 2014]{betancourt2014optimizing}
Betancourt, M., Byrne, S., and Girolami, M. (2014).
\newblock Optimizing the integrator step size for {H}amiltonian {M}onte
  {C}arlo.
\newblock {\em arXiv preprint arXiv:1411.6669}.

\bibitem[Bou-Rabee and Sanz-Serna, 2018]{bou-rabee_sanz-serna_2018}
Bou-Rabee, N. and Sanz-Serna, J.~M. (2018).
\newblock {Geometric integrators and the Hamiltonian Monte Carlo method}.
\newblock {\em Acta Numerica}, 27:113–206.

\bibitem[Burda and Daviet, 2018]{burda2018consumer}
Burda, M. and Daviet, R. (2018).
\newblock {Hamiltonian Sequential Monte Carlo with Application to Consumer
  Choice Behavior}.
\newblock Working Papers tecipa-618, University of Toronto, Department of
  Economics.

\bibitem[Carpenter et~al., 2017]{stan_JSSv076i01}
Carpenter, B., Gelman, A., Hoffman, M., Lee, D., Goodrich, B., Betancourt, M.,
  Brubaker, M., Guo, J., Li, P., and Riddell, A. (2017).
\newblock {Stan: A Probabilistic Programming Language}.
\newblock {\em Journal of Statistical Software, Articles}, 76(1):1--32.

\bibitem[Chopin, 2002]{chopin_ibis}
Chopin, N. (2002).
\newblock A sequential particle filter method for static models.
\newblock {\em Biometrika}, 89(3):539--552.

\bibitem[Chopin and Ridgway, 2017]{chopin2017leave}
Chopin, N. and Ridgway, J. (2017).
\newblock {Leave Pima Indians alone: binary regression as a benchmark for
  Bayesian computation}.
\newblock {\em Statistical Science}, 32(1):64--87.

\bibitem[Chopin et~al., 2013]{chopin2013computational}
Chopin, N., Rousseau, J., and Liseo, B. (2013).
\newblock {Computational aspects of Bayesian spectral density estimation}.
\newblock {\em Journal of Computational and Graphical Statistics},
  22(3):533--557.

\bibitem[Christensen et~al., 2005]{christensen2005scaling}
Christensen, O.~F., Roberts, G.~O., and Rosenthal, J.~S. (2005).
\newblock {Scaling limits for the transient phase of local Metropolis--Hastings
  algorithms}.
\newblock {\em Journal of the Royal Statistical Society: Series B (Statistical
  Methodology)}, 67(2):253--268.

\bibitem[Creutz, 1988]{creutz1988global}
Creutz, M. (1988).
\newblock Global {M}onte {C}arlo algorithms for many-fermion systems.
\newblock {\em Physical Review D}, 38(4):1228.

\bibitem[Daviet, 2018]{daviet2018inference}
Daviet, R. (2018).
\newblock {Inference with Hamiltonian Sequential Monte Carlo Simulators}.
\newblock {\em arXiv preprint arXiv:1812.07978}.

\bibitem[Del~Moral et~al., 2006]{del2006sequential}
Del~Moral, P., Doucet, A., and Jasra, A. (2006).
\newblock Sequential {M}onte {C}arlo samplers.
\newblock {\em Journal of the Royal Statistical Society: Series B (Statistical
  Methodology)}, 68(3):411--436.

\bibitem[Del~Moral et~al., 2007]{delsequential_bayescomp}
Del~Moral, P., Doucet, A., and Jasra, A. (2007).
\newblock Sequential {M}onte {C}arlo for {B}ayesian {C}omputation.
\newblock {\em Bayesian Statistics}, (8):1--34.

\bibitem[Duane et~al., 1987]{duane1987hybrid}
Duane, S., Kennedy, A.~D., Pendleton, B.~J., and Roweth, D. (1987).
\newblock Hybrid {M}onte {C}arlo.
\newblock {\em Physics letters B}, 195(2):216--222.

\bibitem[Fearnhead and Taylor, 2013]{fearnhead2013adaptive}
Fearnhead, P. and Taylor, B.~M. (2013).
\newblock An adaptive sequential {M}onte {C}arlo sampler.
\newblock {\em Bayesian Analysis}, 8(2):411--438.

\bibitem[Friel and Pettitt, 2008]{friel2008marginal}
Friel, N. and Pettitt, A.~N. (2008).
\newblock Marginal likelihood estimation via power posteriors.
\newblock {\em Journal of the Royal Statistical Society: Series B (Statistical
  Methodology)}, 70(3):589--607.

\bibitem[Girolami and Calderhead, 2011]{girolami2011riemann}
Girolami, M. and Calderhead, B. (2011).
\newblock {Riemann manifold Langevin and Hamiltonian Monte Carlo methods}.
\newblock {\em Journal of the Royal Statistical Society: Series B (Statistical
  Methodology)}, 73(2):123--214.

\bibitem[Gorham and Mackey, 2015]{gorham2015measuring}
Gorham, J. and Mackey, L. (2015).
\newblock {Measuring sample quality with Stein's method}.
\newblock In {\em Advances in Neural Information Processing Systems}, pages
  226--234.

\bibitem[Gunawan et~al., 2018]{gunawan2018subsampling}
Gunawan, D., Kohn, R., Quiroz, M., Dang, K.-D., and Tran, M.-N. (2018).
\newblock {Subsampling Sequential Monte Carlo for Static Bayesian Models}.
\newblock {\em arXiv preprint arXiv:1805.03317}.

\bibitem[Hairer et~al., 2003]{hairer2003geometric}
Hairer, E., Lubich, C., and Wanner, G. (2003).
\newblock Geometric numerical integration illustrated by the
  {S}t{\"o}rmer--{V}erlet method.
\newblock {\em Acta numerica}, 12:399--450.

\bibitem[Hairer et~al., 2006]{hairer2006geometric}
Hairer, E., Lubich, C., and Wanner, G. (2006).
\newblock {\em {Geometric numerical integration: structure-preserving
  algorithms for ordinary differential equations}}, volume~31.
\newblock Springer Science \& Business Media.

\bibitem[Hoffman and Gelman, 2014]{hoffman2014no}
Hoffman, M.~D. and Gelman, A. (2014).
\newblock {T}he {N}o-{U}-turn sampler: adaptively setting path lengths in
  {H}amiltonian {M}onte {C}arlo.
\newblock {\em Journal of Machine Learning Research}, 15(1):1593--1623.

\bibitem[Huggins and Roy, 2018]{huggins2015sequential}
Huggins, J.~H. and Roy, D.~M. (2018).
\newblock Sequential monte carlo as approximate sampling: bounds, adaptive
  resampling via $\infty$-ess, and an application to particle gibbs.
\newblock {\em Bernoulli}.

\bibitem[Jasra et~al., 2015]{jasra2015error}
Jasra, A., Paulin, D., and Thiery, A.~H. (2015).
\newblock {Error Bounds for Sequential Monte Carlo Samplers for Multimodal
  Distributions}.
\newblock {\em arXiv preprint arXiv:1509.08775}.

\bibitem[Jasra et~al., 2011]{jasra2011inference}
Jasra, A., Stephens, D.~A., Doucet, A., and Tsagaris, T. (2011).
\newblock {Inference for L{\'e}vy-Driven Stochastic Volatility Models via
  Adaptive Sequential Monte Carlo}.
\newblock {\em Scandinavian Journal of Statistics}, 38(1):1--22.

\bibitem[Kong et~al., 1994]{KongLiuWong}
Kong, A., Liu, J.~S., and Wong, W.~H. (1994).
\newblock Sequential imputation and {B}ayesian missing data problems.
\newblock {\em Journal of the {A}merican statistical association}, 89:278--288.

\bibitem[Kostov, 2016]{kostov2016}
Kostov, S. (2016).
\newblock {Hamiltonian sequential Monte Carlo and normalizing constants}.
\newblock {\em Doctoral thesis, University of Bristol}.

\bibitem[Leimkuhler and Matthews, 2016]{leimkuhler2016molecular}
Leimkuhler, B. and Matthews, C. (2016).
\newblock {\em {Molecular Dynamics}}.
\newblock Springer.

\bibitem[Liu et~al., 2016]{sparsematrix}
Liu, H., Fan, J., and Liao, Y. (2016).
\newblock {An overview of the estimation of large covariance and precision
  matrices}.
\newblock {\em The Econometrics Journal}, 19(1):C1--C32.

\bibitem[Livingstone et~al., 2016]{livingstone2016geometric}
Livingstone, S., Betancourt, M., Byrne, S., and Girolami, M. (2016).
\newblock On the geometric ergodicity of {H}amiltonian {M}onte {C}arlo.
\newblock {\em arXiv preprint arXiv:1601.08057}.

\bibitem[Mangoubi et~al., 2018]{mangoubi2018does}
Mangoubi, O., Pillai, N.~S., and Smith, A. (2018).
\newblock {Does Hamiltonian Monte Carlo mix faster than a random walk on
  multimodal densities?}
\newblock {\em arXiv preprint arXiv:1808.03230}.

\bibitem[Mangoubi and Smith, 2017]{mangoubi2017rapid}
Mangoubi, O. and Smith, A. (2017).
\newblock Rapid mixing of {H}amiltonian {M}onte {C}arlo on strongly log-concave
  distributions.
\newblock {\em arXiv preprint arXiv:1708.07114}.

\bibitem[Minka, 2001]{minka2001expectation}
Minka, T.~P. (2001).
\newblock {Expectation propagation for approximate Bayesian inference}.
\newblock In {\em Proceedings of the Seventeenth conference on Uncertainty in
  artificial intelligence}, pages 362--369. Morgan Kaufmann Publishers Inc.

\bibitem[Mohamed et~al., 2013]{mohamed2013adaptive}
Mohamed, S., de~Freitas, N., and Wang, Z. (2013).
\newblock {Adaptive Hamiltonian and Riemann manifold Monte Carlo samplers}.
\newblock {\em arXiv preprint arXiv:1302.6182}.

\bibitem[Murray et~al., 2016]{murray2016parallel}
Murray, L.~M., Lee, A., and Jacob, P.~E. (2016).
\newblock Parallel resampling in the particle filter.
\newblock {\em Journal of Computational and Graphical Statistics},
  25(3):789--805.

\bibitem[Neal, 1993]{neal1993bayesian}
Neal, R.~M. (1993).
\newblock Bayesian learning via stochastic dynamics.
\newblock In {\em Advances in neural information processing systems}, pages
  475--482.

\bibitem[Neal, 2001]{neal2001annealed}
Neal, R.~M. (2001).
\newblock {Annealed importance sampling}.
\newblock {\em Statistics and computing}, 11(2):125--139.

\bibitem[Neal, 2011]{neal2011mcmc}
Neal, R.~M. (2011).
\newblock {MCMC} using {H}amiltonian dynamics.
\newblock {\em Handbook of Markov Chain Monte Carlo}, 2(11).

\bibitem[Pasarica and Gelman, 2010]{pasarica2010adaptively}
Pasarica, C. and Gelman, A. (2010).
\newblock {Adaptively scaling the Metropolis algorithm using expected squared
  jumped distance}.
\newblock {\em Statistica Sinica}, pages 343--364.

\bibitem[Ridgway, 2016]{ridgway2016computation}
Ridgway, J. (2016).
\newblock {Computation of Gaussian orthant probabilities in high dimension}.
\newblock {\em Statistics and computing}, 26(4):899--916.

\bibitem[Roberts et~al., 1997]{roberts1997weak}
Roberts, G.~O., Gelman, A., and Gilks, W.~R. (1997).
\newblock {Weak convergence and optimal scaling of random walk Metropolis
  algorithms}.
\newblock {\em The annals of applied probability}, 7(1):110--120.

\bibitem[Roberts and Rosenthal, 1998]{roberts1998optimscaling}
Roberts, G.~O. and Rosenthal, J.~S. (1998).
\newblock {Optimal scaling of discrete approximations to Langevin diffusions}.
\newblock {\em Journal of the Royal Statistical Society: Series B (Statistical
  Methodology)}, 60(1):255--268.

\bibitem[Salomone et~al., 2018]{salomone2018unbiased}
Salomone, R., South, L.~F., Drovandi, C.~C., and Kroese, D.~P. (2018).
\newblock {Unbiased and consistent nested sampling via sequential Monte Carlo}.
\newblock {\em arXiv preprint arXiv:1805.03924}.

\bibitem[Sch{\"a}fer and Chopin, 2013]{schafer2013sequential}
Sch{\"a}fer, C. and Chopin, N. (2013).
\newblock {Sequential Monte Carlo on large binary sampling spaces}.
\newblock {\em Statistics and Computing}, pages 1--22.

\bibitem[Schweizer, 2012a]{schweizer2012non}
Schweizer, N. (2012a).
\newblock {Non-asymptotic error bounds for sequential MCMC and stability of
  Feynman-Kac propagators}.
\newblock {\em arXiv preprint arXiv:1204.2382}.

\bibitem[Schweizer, 2012b]{schweizer2012nonmulti}
Schweizer, N. (2012b).
\newblock {Non-asymptotic error bounds for sequential MCMC methods in
  multimodal settings}.
\newblock {\em arXiv preprint arXiv:1205.6733}.

\bibitem[{Sim} et~al., 2012]{SimFilStu}
{Sim}, A., {Filippi}, S., and {Stumpf}, M. P.~H. (2012).
\newblock {Information Geometry and Sequential Monte Carlo}.
\newblock {\em arXiv e-prints}, page arXiv:1212.0764.

\bibitem[Snoek et~al., 2012]{snoek2012practical}
Snoek, J., Larochelle, H., and Adams, R.~P. (2012).
\newblock {Practical Bayesian optimization of machine learning algorithms}.
\newblock In {\em Advances in neural information processing systems}, pages
  2951--2959.

\bibitem[South et~al., 2019]{south2019sequential}
South, L.~F., Pettitt, A.~N., and Drovandi, C.~C. (2019).
\newblock Sequential {M}onte {C}arlo samplers with independent {M}arkov chain
  {M}onte {C}arlo proposals.
\newblock {\em Bayesian Analysis}, 14(3):773--796.

\bibitem[Vats et~al., 2015]{vats2015multivariate}
Vats, D., Flegal, J.~M., and Jones, G.~L. (2015).
\newblock {Multivariate output analysis for Markov chain Monte Carlo}.
\newblock {\em arXiv preprint arXiv:1512.07713}.

\bibitem[Whiteley et~al., 2016]{whiteley2016role}
Whiteley, N., Lee, A., and Heine, K. (2016).
\newblock On the role of interaction in sequential {M}onte {C}arlo algorithms.
\newblock {\em Bernoulli}, 22(1):494--529.

\bibitem[Zhou et~al., 2016]{zhou2016toward}
Zhou, Y., Johansen, A.~M., and Aston, J.~A. (2016).
\newblock {Toward Automatic Model Comparison: An Adaptive Sequential Monte
  Carlo Approach}.
\newblock {\em Journal of Computational and Graphical Statistics},
  25(3):701--726.

\end{thebibliography}
